\documentclass[11pt,a4paper]{article}

\usepackage{hyperref}
\usepackage[top=2cm, bottom=2cm, left=2cm, right=2cm, includefoot]{geometry}

\usepackage{amsthm,amsmath,amssymb,enumerate,mathrsfs}
\usepackage[ruled,linesnumbered,algo2e,vlined]{algorithm2e}
\usepackage{tikz}
\usetikzlibrary{automata,arrows,matrix,fit,topaths,snakes}
\usepackage{cleveref}
\usepackage{wrapfig}
\usepackage[outercaption]{sidecap}
\usepackage{color}

\title{Parameterized DAWGs: \\ efficient constructions and bidirectional pattern searches}

\theoremstyle{definition}
\newtheorem{defn}{Definition}{\bfseries}{\rmfamily}
\newtheorem{example}{Example}{\bfseries}{\rmfamily}
{\bfseries}{\rmfamily}
\newtheorem{remark}{Remark}{\bfseries}{\rmfamily}
\theoremstyle{plain}
\newtheorem{theorem}{Theorem}{\bfseries}{\rmfamily}
\newtheorem{lemma}{Lemma}{\bfseries}{\rmfamily}
\newtheorem{proposition}{Proposition}{\bfseries}{\rmfamily}
\newtheorem{corollary}{Corollary}{\bfseries}{\rmfamily}
\usepackage{authblk}

\newcommand{\occ}{\mathit{occ}}
\newcommand{\rev}[1]{\overline{#1}}
\newcommand{\pr}[1]{\widetilde{#1}} 

\newcommand{\mcal}[1]{\mathcal{#1}}
\newcommand{\mrm}[1]{\mathrm{#1}}

\newcommand{\msf}[1]{\mathsf{#1}}
\newcommand{\mbb}[1]{\mathbb{#1}}
\newcommand{\mtt}[1]{\mathtt{#1}}

\newcommand{\inft}{\protect\scalebox{0.7}[1]{\boldmath$\infty$}}
\newcommand{\calN}{\mcal{N}}

\newcommand{\Prefix}{\msf{Prefix}}
\newcommand{\Factor}{\msf{Factor}}
\newcommand{\Suffix}{\msf{Suffix}}

\newcommand{\PFactor}{\msf{PFactor}}
\newcommand{\PSuffix}{\msf{PSuffix}}

\newcommand{\PSTrie}{\msf{PSTrie}}
\newcommand{\PSTree}{\msf{PSTree}}
\newcommand{\PSAuto}{\msf{PSAuto}}
\newcommand{\PDAWG}{\msf{PDAWG}}
\newcommand{\PPDAWG}{\msf{pPDAWG}}

\newcommand{\parent}{\msf{parent}}

\newcommand{\ssl}{\msf{sl}}

\newcommand{\lrangle}[1]{\langle #1 \rangle}
\newcommand{\lrceil}[1]{\lceil #1 \rceil}
\newcommand{\lrfloor}[1]{\lfloor #1 \rfloor}

\newcommand{\longest}[1]{\lrceil{#1}}
\newcommand{\shortest}[1]{\lrfloor{#1}}
\newcommand{\lec}[2]{[#1]_{#2}^{\mrm{L}}}
\newcommand{\rec}[2]{[#1]_{#2}^{\mrm{R}}}
\newcommand{\llong}[2]{\longest{\lec{#1}{#2}}}
\newcommand{\rlong}[2]{\longest{\rec{#1}{#2}}}
\newcommand{\Zinf}[2]{\langle\!\langle{#1}\rangle\!\rangle_{#2}}
\newcommand{\Update}{\mathsf{update}}
\tikzstyle{edg}=[->,very thick]
\tikzstyle{sfl}=[->,thick,densely dashed,color=blue]
\tikzstyle{dummy}=[draw,-,thick,densely dotted]
\tikzstyle{prelrs}=[draw,thick,color=green,fill=green!12]
\tikzstyle{lrs}=[draw,thick,color=red,fill=red!12]
\tikzstyle{lrslong}=[draw,thick,color=blue,fill=blue!12]

\SetCommentSty{mycommfont}
\SetArgSty{textup}

\begin{document}

\author[1]{Katsuhito~Nakashima}
\author[1]{Noriki~Fujisato}
\author[1]{Diptarama~Hendrian}
\author[2]{Yuto~Nakashima}
\author[1]{Ryo~Yoshinaka}
\author[2,3]{Shunsuke~Inenaga}
\author[4]{Hideo~Bannai}
\author[1]{Ayumi~Shinohara}
\author[2]{Masayuki~Takeda}

\affil[1]{Graduate School of Information Sciences, Tohoku University, Japan}
\affil[2]{Department of Informatics, Kyushu University, Japan}
\affil[3]{PRESTO, Japan Science and Technology Agency, Japan}
\affil[4]{M\&D Data Science Center, Tokyo Medical and Dental University, Japan}

\date{}

\maketitle

\begin{abstract}
	Two strings $x$ and $y$ over $\Sigma \cup \Pi$
	of equal length are said to \emph{parameterized match} (\emph{p-match})
	if there is a renaming bijection $f:\Sigma \cup \Pi \rightarrow \Sigma \cup \Pi$ that is identity on $\Sigma$ and transforms $x$ to $y$ (or vice versa).
	The \emph{p-matching} problem is to look for substrings in a text
	that p-match a given pattern.
	In this paper, we propose
	\emph{parameterized suffix automata} (\emph{p-suffix automata})
	and \emph{parameterized directed acyclic word graphs} (\emph{PDAWGs})
	which are the p-matching versions of suffix automata and DAWGs.
	While suffix automata and DAWGs are equivalent for standard strings,
	we show that p-suffix automata can have $\Theta(n^2)$ nodes and edges
	but PDAWGs have only $O(n)$ nodes and edges,
	where $n$ is the length of an input string.
	We also give an $O(n |\Pi| \log (|\Pi| + |\Sigma|))$-time $O(n)$-space algorithm
	that builds the PDAWG in a left-to-right online manner.
	As a byproduct, it is shown that the \emph{parameterized suffix tree}
	for the reversed string can also be built in the same time and space,
	in a right-to-left online manner.
        This duality also leads us to two further efficient algorithms
        for p-matching:
        Given the parameterized suffix tree for the reversal $\rev{T}$ of the input string $T$,
        one can build the PDAWG of $T$ in $O(n)$ time in an offline manner;
        One can perform \emph{bidirectional} p-matching in $O(m \log (|\Pi|+|\Sigma|) + \occ)$ time using $O(n)$ space,
        where $m$ denotes the pattern length and $\occ$ is the number of pattern occurrences in the text $T$.
\end{abstract}

\section{Introduction}

The \emph{parameterized matching problem} (\emph{p-matching problem})~\cite{PMA}
is a class of pattern matching
where the task is to locate
substrings of a text that have ``the same structure'' as a given pattern.
More formally, we consider a parameterized string (p-string)
over a union of two disjoint alphabets $\Sigma$ and $\Pi$ for static characters
and for parameter characters, respectively.
Two equal length p-strings $x$ and $y$ 
are said to \emph{parameterized match} (\emph{p-match})
if $x$ can be transformed to $y$ (and vice versa)
by a bijection which renames the parameter characters.
The \emph{p-matching problem} is, given a text p-string $T$ and pattern p-string $P$,
to report the occurrences of substrings of $T$ that p-match $P$.
P-matching is well-motivated by plagiarism detection, software maintenance,
RNA structural pattern matching, and so on~\cite{PMA,Shibuya04,MendivelsoP15,MendivelsoTP2020}.

The \emph{parameterized suffix tree} (\emph{p-suffix tree})~\cite{Baker93} is
the fundamental indexing structure for p-matching,
which supports p-matching queries in $O(m \log(|\Pi| + |\Sigma|) + \occ)$ time,
where $m$ is the length of pattern $P$,
and $\occ$ is the number of occurrences to report.
It is known that the p-suffix tree of a text $w$ of length $n$
can be built in $O(n \log (|\Pi| + |\Sigma|))$ time with $O(n)$ space
in an offline manner~\cite{Kosaraju95} and in a \emph{left-to-right online}
manner~\cite{Shibuya04}.
A \emph{randomized} $O(n)$-time left-to-right online construction
algorithm for p-suffix trees is also known~\cite{LeeNP11}.
Indexing p-strings has recently attracted much attention, and 
the p-matching versions of other indexing structures,
such as \emph{parameterized suffix arrays}~\cite{DeguchiHBIT08,TomohiroDBIT09,BealA12a,FujisatoSA2019},
\emph{parameterized BWTs}~\cite{GangulyST17},
and \emph{parameterized position heaps}~\cite{Diptarama2017,Fujisato2018,FujisatoTrie2019},
have also been proposed.

This paper fills in the missing pieces of indexing structures for p-matching,
by proposing the parameterized version of the \emph{directed acyclic word graphs}
(\emph{DAWGs})~\cite{DAWG,Crochemore86},
which we call the \emph{parameterized directed acyclic word graphs} (\emph{PDAWGs}).

For any standard string $T$,
the following three data structures are known to be equivalent:
\begin{enumerate}
\item[(1)] The \emph{suffix automaton} of $T$, which is the minimum DFA that is obtained by merging isomorphic subtrees of the suffix trie of $T$.

\item[(2)] The DAWG, which is the edge-labeled DAG of which each node corresponds to an equivalence class of substrings of $T$ defined by the set of ending positions in $T$.

\item[(3)] The \emph{Weiner-link graph}, which is the DAG consisting of the nodes of the suffix tree of the reversal $\rev{T}$ of $T$ and the reversed suffix links (a.k.a. soft and hard Weiner links).
\end{enumerate}
The equality of (2) and (3) in turn implies symmetry of suffix trees and DAWGs, namely:
\begin{enumerate}
\item[(a)] The suffix links of the DAWG for $T$ form the suffix tree for $\rev{T}$.

\item[(b)] Left-to-right online construction of
the DAWG for $T$ is equivalent to right-to-left online construction of
the suffix tree for $\rev{T}$.
\end{enumerate}

Firstly, we present (somewhat surprising) combinatorial results on the p-matching versions of data structures (1) and (2). We show that the \emph{parameterized suffix automaton} (\emph{p-suffix automaton}), which is obtained by merging isomorphic subtrees of the \emph{parameterized suffix trie} of a p-string $T$ of length $n$, can have $\Theta(n^2)$ nodes and edges in the worst case, while the PDAWG for any p-string has $O(n)$ nodes and edges. On the other hand, the p-matching versions of data structures (2) and (3) are equivalent:
 After introducing the \emph{parameterized Weiner links} on p-suffix trees,
we show that the parameterized Weiner-link graph of the p-suffix tree for $\rev{T}$ is equivalent to the PDAWG for $T$.
As a corollary to this, symmetry (a) also holds: The suffix links of the PDAWG for $T$ form the p-suffix tree for $\rev{T}$.

Secondly, we present algorithmic results on PDAWG construction.
We propose left-to-right \emph{online} construction of PDAWGs that works in
$O(n |\Pi| \log (|\Pi| + |\Sigma|))$ time with $O(n)$ working space.
In addition, as a byproduct of this algorithm, 
we obtain a right-to-left online construction of the p-suffix tree in
$O(n |\Pi| \log (|\Pi| + |\Sigma|))$ time with $O(n)$ space.
This can be seen as the p-matching version of symmetry (b).
The complexities for our online algorithms are valid in the pointer machine,
which is strictly weaker than the word RAM.

Thirdly, we propose an alternative \emph{offline} algorithm which builds
the PDAWG in $O(n)$ time with $O(n)$ working space,
provided that the p-suffix tree of the reversal of the input string is given.
While the proposed offline algorithm itself works in the pointer machine,
there are two different complexities for p-suffix tree construction
in the pointer machine and in the word RAM:
The p-suffix tree can be built offline, 
in $O(n \log (|\Pi|+|\Sigma|))$ time and $O(n)$ space in the pointer machine~\cite{PMA}
and $O(n|\Pi|)$ time and $O(n)$ space in the word RAM with word size $\Omega(\log n)$~\cite{FujisatoSA2019}.
Putting these together, we obtain $O(n \min\{\log(|\Pi|+|\Sigma|), |\Pi|\})$-time $O(n)$-space offline construction of the PDAWG in the word RAM.

We also show that, using the PDAWG for a text string $T$
and the p-suffix tree for its reversal $\rev{T}$,
one can perform \emph{bidirectional} p-matching, i.e., the pattern may grow in both forward and backward directions, in $O(m \log (|\Pi|+|\Sigma|) + \occ)$ time with $O(n)$ space,
where $m$ denotes the pattern length and $\occ$ is the number of pattern occurrences in the text $T$.
To our knowledge, this is the first index that allows bidirectional p-matching
within linear space.

This paper is organized as follows.
After defining basic mathematical notions used in this paper, we will briefly review existing indexing data structures for parameterized strings in Section~\ref{sec:preliminaries}.
We propose three different data structures that can be thought to be the parameterized counterpart of DAWGs in Sections~\ref{sec:psuffix_automata} to~\ref{sec:pdawg_def}:  parameterized suffix automata, pseudo-PDAWGs, and PDAWGs.
The pseudo-PDAWGs are an intermediate data structure which makes the exposition of our pattern matching algorithm with PDAWGs easier to follow.
Section~\ref{sec:duality} discusses the duality between PDAWGs and parameterized suffix trees, together with bidirectional pattern matching
and offline construction algorithms.
Section~\ref{sec:pdawg_construction} presents an algorithm for constructing PDAWGs online.
We conclude the paper in Section~\ref{sec:conclusion}.

A preliminary version of this paper appeared in~\cite{NakashimaFHNYIB20}.
New materials given in this full version are complete proofs
of the lemmas and theorems, introduction of pseudo-PDAWGs data structure,
the bidirectional parameterized pattern matching algorithm,
and more examples and figures.

\section{Preliminaries} \label{sec:preliminaries}
We first introduce definitions and notation used in this paper and then briefly review indexing data structures of parameterized strings.
\subsection{Definitions and notation}

We denote the set of strings over an alphabet $A$ by $A^*$.
For a string $w = xyz \in A^*$, $x$, $y$, and $z$ are called 
\emph{prefix}, \emph{factor}, and \emph{suffix} of $w$, respectively.
The sets of the prefixes, factors, 
and suffixes of a string $w$ are denoted by $\Prefix(w)$, $\Factor(w)$, and $\Suffix(w)$, respectively.
The length of $w$ is denoted by $|w|$ and
 the $i$-th character of $w$ is denoted by $w[i]$ for $1 \leq i \leq |w|$.
The factor of $w$ that begins at position $i$ and ends at position $j$ 
is $w[i:j]$ for $1 \leq i \leq j \leq |w|$.
For convenience,
we abbreviate $w[1:i]$ to $w[{}:i]$ and $w[i:|w|]$ to $w[i:{}]$ for $1 \leq i \leq |w|$.
The empty string is denoted by $\varepsilon$, that is $|\varepsilon|=0$.
Moreover, let $w[i:j] = \varepsilon$ if $i > j$.
The \emph{reverse} $\rev{w}$ of $w \in A^*$ is inductively defined by $\rev{\varepsilon}=\varepsilon$ and $\rev{xa}=a\rev{x}$ for $a \in A$ and $x \in A^*$.

Throughout this paper, we fix two disjoint ordered alphabets $\Sigma$ and $\Pi$.
We call elements of $\Sigma$ \emph{static} characters
and those of $\Pi$ \emph{parameter} characters.
Elements of $\Sigma^{*}$ and $(\Sigma \cup \Pi)^{*}$ are called \emph{static strings} and \emph{parameterized strings} (or \emph{p-strings} for short), respectively.

Given two p-strings $S_1$ and $S_2$ of length $n$,  
$S_1$ and $S_2$ are a \emph{parameterized match} \emph{(p-match)}, denoted by $S_1 \approx S_2$,
if there is a bijection $f$ on $\Sigma \cup \Pi$ such that $f(a) = a$ for any $a \in \Sigma$ and $f(S_1[i]) = S_2[i]$ for all $1 \leq i \leq n$~\cite{PMA}.
We use Kim and Cho's version of p-string encoding~\cite{KimC21}, which replaces $0$ in Baker's prev-encoding~\cite{Baker93} by $\inft$.
Let $\mcal{N} = \mbb{N} \cup \{\inft\} \setminus \{0\}$ where $\mbb{N}$ is the set of non-negative integers.\footnote{%
Without loss of generality, we can assume $\Sigma \cap \calN = \Pi \cap \calN = \emptyset$.
In the case where $\Sigma$ and/or $\Pi$ are integer alphabets, then for instance we can work on the modified alphabets $\Sigma' = \{(0, c) \mid c \in \Sigma\}$, $\Pi' = \{(1, x) \mid x \in \Pi\}$, and $\calN' = \{(2, n) \mid n \in \calN\}$.
}
The \emph{prev-encoding} $\lrangle{S}$ of a p-string $S$ is the string over $\Sigma \cup \calN$ of length $|S|$ defined by
\begin{align*}
	\lrangle{S}[i] = 
	\begin{cases}
    S[i] & \text{if } S[i] \in \Sigma , \\
    \inft          & \text{if } S[i] \in \Pi \text{ and }  S[i] \neq S[j] \text{ for } 1 \le j < i,\\
    i-j  		& \text{if } S[i] = S[j] \in \Pi,\ j<i \text{ and } S[i] \neq S[k] \text{ for any } j < k < i
	\end{cases}
\end{align*}
for $i \in \{1,\dots,|S|\}$.
We call a string $x \in (\Sigma \cup \calN)^*$ a \emph{pv-string} if $x = \lrangle{S}$ for some p-string $S$.
For any p-strings $S_1$ and $S_2$,  $S_1 \approx S_2$ if and only if $\lrangle{S_1}=\lrangle{S_2}$~\cite{PMA}.
For example, given $\Sigma = \{{\tt a, b}\}$ and $\Pi = \{{\tt u, v, x, y}\}$,
$S_1 = {\tt uvvauvb}$ and $S_2 = {\tt xyyaxyb}$ are a p-match 
by $f$ such that $f(\mathtt{u})=\mathtt{x}$ and $f(\mathtt{v})=\mathtt{y}$,
where $\lrangle{S_1} = \lrangle{S_2} = \inft \inft 1{\tt a}43\mtt{b}$.
For a p-string $T$, let $\PFactor(T) = \{\,\lrangle{S} \mid S \in \Factor(T)\,\}$ and $\PSuffix(T) = \{\,\lrangle{S} \mid S \in \Suffix(T)\,\}$ be the sets of prev-encoded factors (pv-factors) and suffixes (pv-suffixes) of $T$, respectively.

Let $S$ be a string of length $n$ over $\Sigma \cup \Pi$,
and $\pi$ denote the number of distinct parameter characters in $S$.
The prev-encoding $\lrangle{S}$ for string $S$
can be computed in $O(n \log \pi) \subseteq O(n \log |\Pi|)$ time using $O(\pi) \subseteq O(n)$ space in an online fashion in the comparison model,
or in $O(n)$ time and space in the word RAM model
when $\Pi$ is an integer alphabet of polynomial size in $n$.

Not every factor of a pv-string is a pv-string.
For example, for $S_1 = {\tt uvvauvb}$ with $\lrangle{S_1}=\inft \inft 1{\tt a}43\mtt{b}$, $\lrangle{S_1}[3:7] =  1{\tt a}43\mtt{b}$ is not a pv-string, but we would like to obtain $\lrangle{S_1[3:7]} = \lrangle{\mtt{vauvb}} = \inft \mtt{a}\inft 3\mtt{b}$ directly from $\inft \inft 1{\tt a}43\mtt{b}$ without ``decoding'' the prev-encoding.
For this end, we extend the notation $\lrangle{\cdot}$ for factors $x\in (\Sigma \cup \calN)^*$ of pv-strings.
The \emph{re-encoding} $\lrangle{x}$ of $x$ is the string of length $|x|$ defined by
\[
 \lrangle{x}[i]= \begin{cases}
	\inft	& \text{if $x[i] \in \calN$ and $x[i] \ge i$,}
\\ 	x[i]	& \text{otherwise.}
 \end{cases}
\]
We then have $\lrangle{\lrangle{S}[i:j]} = \lrangle{S[i:j]}$ for any $i,j$ and $S \in (\Sigma \cup \Pi)^*$.
For example, $\lrangle{\lrangle{S_1}[3:7]} = \lrangle{1\mtt{a}43\mtt{b}} = \inft \mtt{a}\inft 3\mtt{b} = \lrangle{S_1[3:7]}$ for $S_1 = {\tt uvvauvb}$.
We apply $\PFactor$ etc.\ to pv-strings $w$ so that $\PFactor(w)=\{\, \lrangle{x} \mid x \in \Factor(w)\,\}$.
Here, we introduce an alternative definition of the re-encoding using the following notation:
\begin{align*}
 \Zinf{a}{i} = \begin{cases}
	\inft	& \text{if $a \in \calN$ and $a > i$,}
\\ 	a	& \text{otherwise,}
 \end{cases}
\end{align*}
for $a \in \Sigma \cup \calN$ and $i \in \mbb{N}$.
Then, the re-encoding can be defined inductively by $\lrangle{\varepsilon} = \varepsilon$ and $\lrangle{xa}= \lrangle{x}\Zinf{a}{|x|}$ for $x \in (\Sigma \cup \calN)^*$ and  $a \in \Sigma \cup \calN$.

Let $w,x,y \in (\Sigma \cup \calN)^*$.
The set of the \emph{end positions} or \emph{p-occurrences} of $x$ in a pv-string $w$ is defined by $\msf{RPos}_w(x) = \{\, i \in \{0,\dots,|w|\} \mid x = \lrangle{w[i-|x|+1 : i]} \,\}$.
We say $x$ \emph{occurs} in $w$ at $i$ if $i \in \msf{RPos}_w(x)$.
Note that $0 \in \msf{RPos}_w(x)$ iff $x = \varepsilon$.
We write $x \equiv_w^\mrm{R} y$ iff $\msf{RPos}_w(x) = \msf{RPos}_w(y)$
and the equivalence class of $x$ under $\equiv_w^\mrm{R}$ as $[x]_w^\mrm{R}$.
Note that for any $x \notin \PFactor(w)$, which can be a non-pv-string, $\msf{RPos}_w(x) = \emptyset$.
For an equivalent class $u=\rec{x}{w}$, we may write $\msf{RPos}_w(u)$ to mean $\msf{RPos}_w(x)$ for $x \in u$.
For a finite nonempty set $X$ of strings which has no distinct elements of equal length, the shortest and longest elements of $X$ are denoted by $\shortest{X}$ and $\longest{X}$, respectively.
\begin{example}
Let $w = \lrangle{\mtt{x a x a y a}} =  \inft \mtt{a} 2 \mtt{a} \inft \mtt{a}$ where $\Sigma = \{\mtt{a}\}$ and $\Pi = \{\mtt{x},\mtt{y}\}$.
Then, $\msf{RPos}_w(\mtt{a})=\msf{RPos}_w(\inft\mtt{a})=\{2,4,6\}$ and $\rec{\mtt{a}}{w}=\{\mtt{a},\inft\mtt{a}\}$.
On the other hand, $\msf{RPos}_w(2\mtt{a})=\emptyset$ and $\rec{2\mtt{a}}{w}=(\Sigma \cup \calN)^* \setminus \PFactor(w)$.
\end{example}

The \emph{parameterized pattern matching problem} is to enumerate all the end positions of $\lrangle{P}$ in $\lrangle{T}$, i.e., elements of $\msf{RPos}_{\lrangle{T}}(\lrangle{P})$, for given two p-strings $T$ and $P$.
The weaker version of the problem is to decide whether $\msf{RPos}_{\lrangle{T}}(\lrangle{P}) \neq \emptyset$.

The basic properties on end positions of factors in static strings presented in~\cite{DAWG} also hold for pv-strings.
\begin{lemma}\label{lem:basic}
	Let $x$, $y$ and $w$ be pv-strings.
	If\/ $\msf{RPos}_w(x) \cap \msf{RPos}_w(y) \neq \emptyset$, then either $x \in \PSuffix(y)$ or $y \in \PSuffix(x)$.
	If $x \in \PSuffix(y)$, then $\msf{RPos}_w(y) \subseteq \msf{RPos}_w(x)$.
	For any $x \in \PFactor(w)$, there is $k \in \mbb{N}$ such that
	\(
		\rec{x}{w} = \{\, \lrangle{y[i:{}]} \mid 1 \le i \le k \,\}
	\) where $y=\longest{\rec{x}{w}}$.
\end{lemma}
We will use the above lemma implicitly in arguments in this paper.

We define notions symmetric to $\msf{RPos}$, $\equiv^{\mrm{R}}$, and $[\cdot]^\mrm{R}$.
The start position set of $x$ in a pv-string $w$ is $\msf{LPos}_w(x) = \{\,i \in \{0,\dots,|w|\} \mid \lrangle{w[i:i+|x|-1]}=x\,\}$.
We write $x \equiv_w^\mrm{L} y$ iff $\msf{LPos}_w(x) = \msf{LPos}_w(y)$.
The equivalence class of $x$ under $\equiv_w^\mrm{L}$ is denoted by $[x]_w^\mrm{L}$.

\subsection{Existing indexing structures for parameterized strings}
A basic indexing structure of a p-string is a \emph{parameterized suffix trie}.
The parameterized suffix trie $\PSTrie(T)$ is the trie for $\PSuffix(T)$.
That is, $\PSTrie(T)$ is an edge-labeled tree $(V,E)$ whose node set is $V=\PFactor(T)$ and edge set is $E=\{\,(x,a,xa) \in V \times (\Sigma \cup \calN) \times V \,\}$.
We remark that each edge of $\PSTrie(T)$ is labeled by a single symbol from the prev-encoding $\langle T \rangle$, and that the out-going edge labels of each node are mutually distinct.
An example of $\PSTrie(T)$ can be found in Figure~\ref{fig:pSAoto_pDAWG}(a).
Like the standard suffix tries for static strings, the size of $\PSTrie(T)$ can be $\Theta(|T|^2)$.
Obviously we can check whether $T$ has a substring that p-matches $P$ of length $m$ in $O(m \log(|\Pi|+|\Sigma|))$ time using $\PSTrie(T)$,
assuming that finding the edge to traverse for a given character takes $O(\log(|\Pi|+|\Sigma|))$ time by, e.g., using balanced trees.
We use the same assumption on other indexing structures considered in this paper.

A more compact representation of the suffix sets of p-strings is \emph{parameterized suffix trees}~\cite{Baker93}.
The parameterized suffix tree $\PSTree(T)$ of a p-string $T$ is the path-compacted (or Patricia) tree for $\PSuffix(T)$.
That is, $\PSTree(T)$ is an edge-labeled tree $(V,E)$ of $T$ where for $w=\lrangle{T}$,
\begin{align*}
	V &= \{\, \llong{x}{w} \mid x \in \PFactor(T) \,\},
\\	E &= \{\, (x,y,xy) \in V \times (\Sigma \cup \calN)^+ \times V  \mid xy = \llong{xa}{w} \in V \text{ for some } a \in \Sigma \cup \mcal{N}\,\}.
\end{align*}
We remark that each edge of $\PSTree(T)$ is labeled by an element of $\Factor(\PSuffix(T)) \setminus \{\varepsilon\}$, and that the labels of the out-going edges of each node begin with mutually distinct symbols.
An example of $\PSTree(T)$ can be found in Figure~\ref{fig:pSAoto_pDAWG}(b).

To store $\PSTree(T)$ in linear space, in an actual implementation, the label $y$ of an edge $(x,y,xy)$ is represented by two integers $i$ and $j$ such that $y=\lrangle{T[i-|x|:j]}[|x|+1:{}]$, where $\lrangle{T[i-|x|:j]}=xy$.
In other words, $y$ corresponds to $T[i:j]$ but the prev-encoding is given relative to $T[i-|x|:j]$. 
The value $|x|$ is stored in the node $x$, though it is possible to calculate $|x|$ by reading edge labels from the root to $x$.

\section{Parameterized suffix automata}\label{sec:psuffix_automata}
Recall that the DAWG for a static string $w \in \Sigma^*$ is isomorphic to a minimal deterministic finite automaton that accepts all the suffixes of $w$, which can be obtained by merging isomorphic subtrees of the suffix tries.
A static string $x$ occurs in $w$ if and only if the automaton has a state that one can reach by reading $x$. 
One natural idea to define the parameterized counterpart of DAWGs, which we actually do not take, may be to have minimal deterministic finite automata for prev-encoded suffixes.
This is equivalent to merging isomorphic subtrees of parameterized suffix tries.
More formally, letting $w=\lrangle{T}$ and $[x]_w^\mrm{N} = \{\, y \in \PFactor(w) \mid xz \in \PSuffix(w) \Leftrightarrow yz \in \PSuffix(w) $ for all $z \in (\Sigma \cup \mcal{N})^* \,\}$ (Nerode equivalence class), $\PSAuto(T)$ is defined as a directed acyclic graph $(V,E)$ with
\begin{align*}
	V &= \{\, [x]_w^\mrm{N} \subseteq \PFactor(w) \mid x \in \PFactor(w)\,\},
\\	E &= \{\, ([x]_w^\mrm{N},a,[xa]_w^\mrm{N}) \in V \times (\Sigma \cup \mcal{N}) \times V \mid xa \in \PFactor(w) \,\}\,,
\end{align*}
where the initial state is $[\varepsilon]_w^\mrm{N}$ and the final states are $[x]_w^\mrm{N}$ for $x \in \PSuffix(w)$.
We then have $x \in \PFactor(w)$ if and only if one can reach some state in the automaton by reading $x$. 
Figure~\ref{fig:pSAoto_pDAWG}(c) shows an example of a parameterized suffix automaton, where we do not distinguish final states and other states.
However, differently from the case of static strings, the size of $\PSAuto(T)$ can be $\Theta(|T|^2)$ for $T \in (\Sigma \cup \Pi)^*$.
\begin{figure}[t]
	\centering
	\newcommand{\gate}[2]{\inft\scriptsize/\textcolor{red}{$[{#1}\!:\!{#2}]$}}
	\newcommand{\nnm}[1]{\scriptsize\textsf{#1}}
	\begin{tikzpicture}[xscale=0.65,thick,shape=circle,inner sep=0pt,minimum size=4mm]
	\draw (0,-5.75) node {\small(a)};
	\node[] (dummy) at (0,1) {};
	\node[draw] (e) at (0,0) {\nnm{0}};
	\node[draw] (a) at (-1,-1.25) {\nnm{6}};
	\node[draw] (a0) at (-1,-2.25) {\nnm{7}};
	\node[draw] (a0a) at (-1,-3.25) {\nnm{8}};
	\node[draw] (a0a0) at (-1,-4.25) {\nnm{9}};
	\node[draw] (0) at (0,-1) {\nnm{1}};
	\node[draw] (0a) at (0,-2) {\nnm{2}};
	\node[draw] (0a0) at (1,-3.5) {\nnm{10}};
	\node[draw] (0a2) at (0,-3) {\nnm{3}};
	\node[draw] (0a2a) at (0,-4) {\nnm{4}};
	\node[draw] (0a2a0) at (0,-5) {\nnm{5}};
	\draw[edg] (dummy) -- (e);
	\draw[edg,out=180,in=90] (e) to node[left] {$\mtt{a}$} (a);
	\draw[edg] (a) to node[left] {$\inft$} (a0);
	\draw[edg] (a0) to node[left] {$\mtt{a}$} (a0a);
	\draw[edg] (a0a) to node[left] {$\inft$} (a0a0);
	\draw[edg] (e) to node[right] {$\inft$} (0);
	\draw[edg] (0) to node[right] {$\mtt{a}$} (0a);
	\draw[edg,out=0,in=90] (0a) to node[right] {$\inft$} (0a0);
	\draw[edg] (0a) to node[right] {$2$} (0a2);
	\draw[edg] (0a2) to node[right] {$\mtt{a}$} (0a2a);
	\draw[edg] (0a2a) to node[right] {$\inft$} (0a2a0);
	\end{tikzpicture}
	\begin{tikzpicture}[xscale=0.6,thick,shape=circle,inner sep=0pt,minimum size=4mm]
	\draw (0,-5.75) node {\small(b)};
	\node[] (dummy) at (0,1) {};
	\node[draw] (e) at (0,0) {\nnm{0}};
	\node[draw] (a0) at (-1,-2.25) {\nnm{7}};
	\node[draw] (a0a0) at (-1,-4.25) {\nnm{9}};
	\node[draw] (0) at (0,-1) {\nnm{1}};
	\node[draw] (0a) at (0,-2) {\nnm{2}};
	\node[draw] (0a0) at (1,-3.5) {\nnm{10}};
	\node[draw] (0a2a0) at (0,-5) {\nnm{5}};
	\draw[edg] (dummy) -- (e);
	\draw[edg,out=180,in=90] (e) to node[left] {$\mtt{a}\inft$} (a0);
	\draw[edg] (a0) to node[left] {$\mtt{a}\inft$} (a0a0);
	\draw[edg] (e) to node[right] {$\inft$} (0);
	\draw[edg] (0) to node[right] {$\mtt{a}$} (0a);
	\draw[edg,out=0,in=90] (0a) to node[right] {$\inft$} (0a0);
	\draw[edg] (0a) to node[right,near end] {$2\mtt{a}\inft$} (0a2a0);
	\end{tikzpicture}
	\begin{tikzpicture}[xscale=0.65,thick,shape=circle,inner sep=0pt,minimum size=4mm]
	\draw (0,-5.75) node {\small(c)};
	\node[] (dummy) at (0,1) {};
	\node[draw] (e) at (0,0) {\nnm{0}};
	\node[draw] (a) at (-1,-1.25) {\nnm{6}};
	\node[draw] (a0) at (-1,-2.75) {\nnm{7}};
	\node[draw] (0) at (0,-1) {\nnm{1}};
	\node[draw] (0a) at (0,-2) {\nnm{2}};
	\node[draw] (0a2) at (0,-3) {\nnm{3}};
	\node[draw] (0a2a) at (0,-4) {\nnm{4},\nnm{8}};
	\node[draw] (0a2a0) at (0,-5) {};
	\node (0a2a0Label1) at (0.02,-4.95) {\nnm{\tiny{}5},\nnm{\tiny{}9},};
	\node (0a2a0Label2) at (0,-5.1) {\nnm{\tiny{}10}};
	\draw[edg] (dummy) -- (e);
	\draw[edg,out=190,in=90] (e) to node[left] {$\mtt{a}$} (a);
	\draw[edg] (a) to node[left] {$\inft$} (a0);
	\draw[edg,out=270,in=170] (a0) to node[left] {$\mtt{a}$} (0a2a);
	\draw[edg] (e) to node[right] {$\inft$} (0);
	\draw[edg] (0) to node[right] {$\mtt{a}$} (0a);
	\draw[edg,out=340,in=20] (0a) to node[right] {$\inft$} (0a2a0);
	\draw[edg] (0a) to node[right] {$2$} (0a2);
	\draw[edg] (0a2) to node[right] {$\mtt{a}$} (0a2a);
	\draw[edg] (0a2a) to node[right] {$\inft$} (0a2a0);
	\end{tikzpicture}
	\begin{tikzpicture}[xscale=0.75,thick,shape=circle,inner sep=0pt,minimum size=4mm]
	\draw (0,-5.75) node {\small(d)};
	\node[] (dummy) at (0,1) {};
	\node[draw] (e) at (0,0) {\nnm{0}};
	\node[draw] (a0) at (-1,-3) {\nnm{7}};
	\node[draw] (0) at (0,-1) {\nnm{1}};
	\node[draw] (0a) at (0,-2) {\nnm{2},\nnm{6}};
	\node[draw] (0a2) at (0,-3) {\nnm{3}};
	\node[draw] (0a2a) at (0,-4) {\nnm{4},\nnm{8}};
	\node[draw] (0a2a0) at (0,-5) {};
	\node (0a2a0Label1) at (0.02,-4.95) {\tiny\nnm{\tiny{}5},\nnm{\tiny{}9},};
	\node (0a2a0Label2) at (0,-5.1) {\nnm{\tiny{}10}};
	\draw[edg] (dummy) -- (e);
	\draw[edg,out=190,in=160] (e) to node[left] {$\mtt{a}$} (0a);
	\draw[edg,out=200,in=80] (0a) to (a0);
		\draw (-0.64,-2.5) node{\gate{1}{1}};
	\draw[edg,out=280,in=160] (a0) to node[below] {$\mtt{a}$} (0a2a);
	\draw[edg] (e) to node[right] {\gate{0}{0}} (0);
	\draw[edg] (0) to node[right] {$\mtt{a}$} (0a);
	\draw[edg] (0a) to [out=350,in=90] (1.25,-3) to (1.25,-3.5) node[right=-11pt] {\gate{2}{2}} to (1.25,-4.25) to [out=270,in=0] (0a2a0);
	\draw[edg] (0a) to node[right] {$2$} (0a2);
	\draw[edg] (0a2) to node[right] {$\mtt{a}$} (0a2a);
	\draw[edg] (0a2a) to node[right=-11pt] {\gate{3}{4}} (0a2a0);
	\draw[sfl,out=160,in=270] (0a2a0) to (a0);
	\draw[sfl,out=130,in=180] (a0) to (0);
	\draw[sfl,out=150,in=210] (0) to (e);
	\draw[sfl,out=40,in=330] (0a2a) to (0a);
	\draw[sfl] (0a) to [out=15,in=270] (1.35,-0.5) to [out=90,in=0] (e);
	\draw[sfl,out=180,in=0] (0a2) to (a0);
	\end{tikzpicture}
	\begin{tikzpicture}[xscale=0.75,thick,shape=circle,inner sep=0pt,minimum size=4mm]
	\draw (0,-5.75) node {\small(e)};
	\node[] (dummy) at (0,1) {};
	\node[draw] (e) at (0,0) {\nnm{0}};
	\node[draw] (a0) at (-1,-3) {\nnm{7}};
	\node[draw] (0) at (0,-1) {\nnm{1}};
	\node[draw] (0a) at (0,-2) {\nnm{2},\nnm{6}};
	\node[draw] (0a2) at (0,-3) {\nnm{3}};
	\node[draw] (0a2a) at (0,-4) {\nnm{4},\nnm{8}};
	\node[draw] (0a2a0) at (0,-5) {};
	\node (0a2a0Label1) at (0.02,-4.95) {\tiny\nnm{\tiny{}5},\nnm{\tiny{}9},};
	\node (0a2a0Label2) at (0,-5.1) {\nnm{\tiny 10}};
	\draw[edg] (dummy) -- (e);
	\draw[edg,out=190,in=160] (e) to node[left] {$\mtt{a}$} (0a);
	\draw[edg,out=280,in=160] (a0) to node[below] {$\mtt{a}$} (0a2a);
	\draw[edg] (e) to node[right] {$\inft$} (0);
	\draw[edg] (0) to node[right] {$\mtt{a}$} (0a);
	\draw[edg,out=340,in=20] (0a) to node[right] {$\inft$} (0a2a0);
	\draw[edg] (0a) to node[right] {$2$} (0a2);
	\draw[edg] (0a2) to node[right] {$\mtt{a}$} (0a2a);
	\draw[edg] (0a2a) to node[right] {$\inft$} (0a2a0);
	\draw[sfl,out=160,in=270] (0a2a0) to (a0);
	\draw[sfl,out=90,in=215] (a0) to (0);
	\draw[sfl,out=150,in=210] (0) to (e);
	\draw[sfl,out=40,in=320] (0a2a) to (0a);
	\draw[sfl,out=0,in=0] (0a) to (e);
	\draw[sfl,out=180,in=0] (0a2) to (a0);
	\end{tikzpicture}
	\caption{\label{fig:pSAoto_pDAWG}
		(a) The parameterized suffix trie $\PSTrie(T)$,
		(b) the parameterized suffix tree $\PSTree(T)$,
		(c) the parameterized suffix automaton $\PSAuto(T)$,
		(d) the pseudo PDAWG $\PPDAWG(T)$,
		 and (e) the PDAWG $\PDAWG(T)$ for $T=\mtt{xaxay}$ over $\Sigma= \{\mtt{a}\}$ and $\Pi= \{\mtt{x},\mtt{y}\}$, for which $\lrangle{T}= w = \inft \mtt{a} 2 \mtt{a} \inft$.
		Solid and broken arrows represent the edges and suffix links, respectively.
		Gate intervals of $\inft$-edges of $\PPDAWG(T)$ are shown with red letters. 
		Some nodes of $\PDAWG(T)$ cannot be reached by following edges from the source node.
		The numbers in nodes illustrate how nodes in $\PSTrie(T)$ are skipped in $\PSTree(T)$ or merged in the other structures. 
	}
\end{figure}
\begin{proposition}\label{prop:psauto}
	The size of\/ $\PSAuto(T)$ is $\Theta(|T|^2)$.
\end{proposition}
\begin{proof}
	Let $T_k = \mathtt{x}_1 \mathtt{a}_1 \dots \mathtt{x}_k \mathtt{a}_k \mathtt{x}_1 \mathtt{a}_1 \dots \mathtt{x}_k \mathtt{a}_k $ be a p-string over $\Sigma_k = \{\mtt{a}_1,\dots,\mtt{a}_k\}$ and $\Pi_k=\{\mtt{x}_1, \dots \mtt{x}_k\}$, where $|T_k|=4k$.
	For  $1 \le i < j \le k$, we have $y_{j,i} = \inft \mtt{a}_j \dots \inft \mtt{a}_k \inft \mtt{a}_1 \dots \inft \mtt{a}_i = \lrangle{T_k[2j-1:2k+2i]}\in \PFactor(T_k)$.
	We show that we reach different nodes by reading $y_{j,i}$ and $y_{j',i'}$ unless $i=i'$ and $j=j'$.
	If $i \neq i'$ or $j \neq j'$,  $\inft \mtt{a}_{i+1} \dots \inft \mtt{a}_{j-1} (2k) \mtt{a}_j$ can follow $y_{j,i}$ but not $y_{j',i'}$ to form an element of $\PFactor(T_k)$.
	Therefore, $\PSAuto(T_k)$ must have at least $k(k-1)/2 \in \Theta(k^2)$ nodes.
\end{proof}
For example, for $k=3$, $i=i'=1$, $j=2$, and $j'=3$, we have $y_{2,1} = \lrangle{\mtt{x}_2 \mtt{a}_2 \mtt{x}_3 \mtt{a}_3 \mtt{x}_1 \mtt{a}_1} = \inft \mtt{a}_2 \inft \mtt{a}_3 \inft \mtt{a}_1$ and $y_{3,1} = \lrangle{\mtt{x}_3 \mtt{a}_3 \mtt{x}_1 \mtt{a}_1} = \inft \mtt{a}_3 \inft \mtt{a}_1$.
Then, $z = 6 \mtt{a}_2$ can follow $y_{2,1}$, i.e., $y_{2,1}z = \lrangle{\mtt{x}_2 \mtt{a}_2 \mtt{x}_3 \mtt{a}_3 \mtt{x}_1 \mtt{a}_1 \mtt{x}_2 \mtt{a}_2} \in \PFactor(T_3)$.
However, $y_{3,1} z$ is not a pv-string, so not in $\PFactor(T_3)$.
We conclude $[y_{2,1}]_{\lrangle{T_3}}^\mrm{N} \not\equiv [y_{3,1}]_{\lrangle{T_3}}^\mrm{N}$.
We remark that Proposition~\ref{prop:psauto} holds under binary alphabets, too, which can be shown by the standard binary encoding technique.

We will seek better ideas to define parameterized DAWGs in the following sections.

\section{Pseudo parameterized directed acyclic word graphs} \label{sec:pdawg_pseudo}
Another way to define a parameterized counterpart of DAWGs for static strings may be to merge nodes with the same end position sets in parameterized suffix tries.
This approach and the one in the previous section result in the same structures for static strings, but it is not the case for p-strings.
We call such a structure for a p-string $T$ a \emph{pseudo-PDAWG} and denote it by $\PPDAWG(T)$,
which can formally be defined as a direct acyclic graph $(V,E)$ where for $w=\lrangle{T}$
\begin{align*}
	V &= \{\, \rec{x}{w} \mid x \in \PFactor(w)\,\},
\\	E &= \{\, (\rec{x}{w},a,\rec{xa}{w}) \in V \times (\Sigma \cup \calN) \times V \mid xa \in \PFactor(w)\,\}\,.
\end{align*}
The nodes $[\varepsilon]^\mrm{R}_w$ and $[w]^\mrm{R}_w$ are called the \emph{source} and the \emph{sink}, respectively.
An edge $(u,a,v) \in E$ is called the \emph{$a$-edge} of $u$.
In this section, we fix $w = \lrangle{T}$.
Nodes of a parameterized suffix trie merged in the parameterized suffix automaton are also merged in the pseudo-PDAWG but not vice versa.
Therefore, pseudo-PDAWGs can be smaller than parameterized suffix automata.
In fact, as we will see later (Theorem~\ref{thm:pdawg_size}), the number of nodes of a pseudo-PDAWG is linearly bounded in $|T|$.
If $T$ has no parameter characters, our pseudo-PDAWGs coincide with DAWGs for static strings~\cite{DAWG}.

However, this idea results in an apparent conflict for p-strings.
Figure~\ref{fig:pSAoto_pDAWG}(d) shows the pseudo-PDAWG $\PPDAWG(T)$ obtained from $\PSTrie(T)$ of Figure~\ref{fig:pSAoto_pDAWG}(a) for $T=\mtt{xaxay}$.
Concerning the two nodes $\mtt{a}$ and $\inft\mtt{a}$ in $\PSTrie(T)$, we have $\msf{RPos}_w(\mtt{a})=\msf{RPos}_w(\inft\mtt{a})=\{2,4\}$, so they shall be merged in $\PPDAWG(T)$.
However, the subtrees rooted at $\mtt{a}$ and $\inft\mtt{a}$ in $\PSTrie(T)$ have different shapes.
The $\inft$-edges of those two nodes point at nodes $\mtt{a}\inft$ and $\inft\mtt{a}\inft$, with $\msf{RPos}_w(\mtt{a}\inft) \neq \msf{RPos}_w(\inft\mtt{a}\inft)$, which shall not be merged.
As a result of merging $\mtt{a}$ and $\inft\mtt{a}$, the obtained node in $\PPDAWG(T)$ has two edges labeled with the same character $\inft$ pointing at different nodes.
Consequently, $\PPDAWG(T)$ has got a path labeled with $\inft\mtt{a}\inft\mtt{a}\inft$, which is not in $\PFactor(T)$.
This apparently obstructs pattern matching over $\PPDAWG(T)$.
Below we will explain how to resolve the problem and will show nice properties of pseudo-PDAWGs as indexing structures.
Lemma~\ref{lem:ppdawg1} shows that if a node $u$ in a pseudo-PDAWG has two or more edges labeled with the same character, the character must be $\inft$.
We will determinize $\inft$-edges by giving them mutually exclusive ``gate intervals'' so that one can follow an $\inft$-edge only when the string read so far has a length in the interval, based on Lemma~\ref{lem:ppdawg2}.
In Figure~\ref{fig:pSAoto_pDAWG}(d), the gate intervals are shown in red beside those $\inft$-edges.
One can follow an $\inft$-edge with a gate interval $[i:j]$ only when $\inft$ follows a prefix of length between $i$ and $j$. 
This prevents one to follow the path labeled with $\inft\mtt{a}\inft\mtt{a}\inft$ in Figure~\ref{fig:pSAoto_pDAWG}(d).

\begin{lemma}\label{lem:ppdawg1}
	Given pv-strings $xa$, $ya$, and $w$ with $a \in \Sigma \cup \mcal{N}$,
	if $x \equiv_w^\mrm{R} y$ and $xa \not\equiv_w^\mrm{R} ya$, then $a=\inft$.
\end{lemma}
\begin{proof}
	Suppose that $a \in \Sigma$ and $x \equiv_w^\mrm{R} y$.
	If $i \in \msf{RPos}_w(xa)$, then $i-1 \in \msf{RPos}_w(x) = \msf{RPos}_w(y)$ and $w[i] = a$.
	This means $i \in \msf{RPos}_w(ya)$. 
	Therefore, $\msf{RPos}_w(xa) \subseteq \msf{RPos}_w(ya)$ and symmetrically we can show $\msf{RPos}_w(ya) \subseteq \msf{RPos}_w(xa)$.
	That is, $xa \equiv_w^\mrm{R} ya$.

	Suppose $a \in \calN \setminus \{\inft\}$ and $x \equiv_w^\mrm{R} y$.
	The facts that $xa$ and $ya$ are pv-strings and $a \in \mbb{N}$ imply $|x|,|y| \ge a$.
	If $i \in \msf{RPos}_w(xa)$, then ${w}[i]=a$.
	Since $i-1 \in \msf{RPos}_w(x)=\msf{RPos}_w(y)$, we have $i \in \msf{RPos}_w(ya)$.
	Therefore, $\msf{RPos}_w(xa) = \msf{RPos}_w(ya)$.

	Hence, only when $a=\inft$, it is possible that $x \equiv_w^\mrm{R} y$ and $xa \not\equiv_w^\mrm{R} ya$.
\end{proof}
The following lemma means that $\inft$-edges of a node have respective gate intervals $[i:j]$ according to which one can choose the valid one to follow.
\begin{lemma}\label{lem:ppdawg2}
	Each $\inft$-edge $(u,\inft,v)$ of\/ $\PPDAWG(T)$ admits two integers $i$ and $j$ such that for any $x \in u$, we have $x\inft \in v$ if and only if\/ $i \le |x| \le j$.
\end{lemma}
\begin{proof}
	Let $i=|y|$ and $j=|z|$ for the shortest $y$ and longest $z$ such that $y,z \in u$ and $y\inft,z\inft \in v$.
	Then, the ``only if'' direction is obvious.
	On the other hand, for every $x$ such that $x \in u$ and $i \le |x| \le j$, we have $y \in \PSuffix(x)$ and $x \in \PSuffix(z)$ by Lemma~\ref{lem:basic}.
	This implies, again by Lemma~\ref{lem:basic}, $x\inft \in v$.
\end{proof}
Now, we enhance $\inft$-edges of pseudo-PDAWGs with the gate intervals $[i:j]$ given in Lemma~\ref{lem:ppdawg2}.
Suppose we have reached a node $u$ by reading a pv-string $x$.
If the next character $a$ is not $\inft$, we simply follow the $a$-edge of $u$.
If $a=\inft$, we follow the $\inft$-edge with interval $[i:j]$ such that $i \le |x| \le j$.
If $u$ has no such edge, it means that $xa \notin \PFactor(T)$.
We remark that $\PPDAWG(T)$ in Figure~\ref{fig:pSAoto_pDAWG}(d) has a path $\mtt{a}2$ which does not exist in $\PSTrie(T)$, but it is harmless, since the non-pv-string $\mtt{a}2$ can be obtained from no input pattern $P$.
\begin{proposition}
	Let $w = \lrangle{T}$.
	One can reach a node $u$ from $\rec{\varepsilon}{w}$ by reading $x$ in $\PPDAWG(T)$ if and only if\/ $u=\rec{x}{w}$.
\end{proposition}
\begin{proof}
	We show the proposition by induction on $|x|$.
	If $x=\varepsilon$, the conclusion is trivial.
	Suppose we have reached $u=\rec{x}{w}$ by reading $x$ and the next character is $a \in \Sigma \cup \mcal{N} \setminus \{\inft\}$.
	If\/ $xa \in \PFactor(T)$, Lemma~\ref{lem:ppdawg1} implies that $u$ has only one $a$-edge $(u,a,v)$, for which $v=\rec{xa}{w}$ holds.
	If\/ $xa \notin \PFactor(T)$, then $ya \notin \PFactor(T)$ for all $y \in u$, which means that $u$ has no $a$-edge by the definition of $E$.
	Suppose the next character is $a=\inft$.
	Lemma~\ref{lem:ppdawg2} implies that $u$ has only one $\inft$-edge $(u,\inft,v)$ with a length condition $[i:j]$ satisfying $i \le |x| \le j$, for which $v=\rec{x\inft}{w}$,
	 if and only if\/ $xa \in \PFactor(T)$.
\end{proof}
Therefore, one can decide whether $\lrangle{P} \in \PFactor(T)$ in time $O(|P| \log (|\Pi|+|\Sigma|))$ using $\PPDAWG(T)$.

\begin{algorithm2e}[t]
	\newcommand{\nonl}{\renewcommand{\nl}{\let\nl\oldnl}}
	\caption{Parameterized pattern matching algorithm based on $\PPDAWG(T)$\label{alg:ppdawgmatching}}
	\SetVlineSkip{0.5mm}
	Let $p \leftarrow \lrangle{P}$\;
	Let $u \leftarrow [\varepsilon]^\mrm{R}_{\lrangle{T}}$\; 
	\For{$i = 1$ \textup{\textbf{to}} $|p|$}{%
		\lIf{$p[i] \neq \inft$ and $(u,p[i],v) \in E$}{Let $u \leftarrow v$}
		\lElseIf{$u$ has an $\inft$-edge $(u,\inft,v) \in E$ whose gate interval $[j:k]$ satisfies $j \le i-1 \le k$}{Let $u \leftarrow v$}
		\lElse{\textbf{return} $\mathsf{False}$}
	}
	Traverse the reversed suffix link tree and \\ \nonl \quad \textbf{output} $\ell_v$ for all descendants $v$ of $u$\;
\end{algorithm2e}

In order to find all end positions of substrings of the text which p-matches with an input p-string, we further augment pseudo-PDAWGs, in the way analogous to the classical enhancement of DAWGs.
One obvious idea might be to explicitly record $\msf{RPos}_w(u)$ in each node $u \in V$ so that all the p-occurrences of $x \in u$ can be found at the reached node, but it makes the data structure size non-linear.
Instead, we assign the smallest number $\ell_u$ in $\msf{RPos}_w(u)$ to each node $u$, i.e., the end position of the left most p-occurrence of $x \in u$ in $w$.
To find other p-occurrences, \emph{suffix links} are useful, which are defined as
\[
	F = \{\, (u,v) \in (V \setminus \{\{\varepsilon\}\}) \times V \mid v = \rec{\lrangle{x[2:{}]}}{w} \text{ for } x = \shortest{u}\,\}
\,.\]
In other words, for two nodes $u,v \in V$, we have $(u,v) \in F$ if and only if $\longest{v}$ is obtained by removing the first character of $\shortest{u}$.
Thus, every pv-suffix of $x \in u$ belongs to some $v \in V$ which can be reached from $u$ by following the suffix links.
Consequently, the reverse of suffix links $\overline{F}=\{\,(v,u) \mid (u,v) \in F \,\}$ form a rooted tree, where $\rec{\varepsilon}{w}$ is the root and $v$ is a child of $u$ if $(u,v) \in \overline{F}$.
We call $(V,\overline{F})$ \emph{the reversed suffix link tree}.
We will discuss in Section~\ref{sec:duality} that $(V,\overline{F})$ is isomorphic to the parameterized suffix tree of $\overline{T}$, from which it is obvious that one can find all p-occurrences of $x$ in $w$ by visiting all the descendants of the reached node $\rec{x}{w}$ in $(V,\overline{F})$.
Here, independently of the duality arguments, we justify the p-matching procedure with reversed suffix links.
\begin{lemma}\label{lem:pdawg_matching}
	It holds that $k \in \msf{RPos}_w(v)$ if and only if $k = \ell_u$ for some (not necessarily proper) descendant $u$ of $v$ in $(V,\overline{F})$.
\end{lemma}
\begin{proof}
	If $u$ is a child of $v$, then $\msf{RPos}_w(u) \subsetneq \msf{RPos}_w(v)$.
	This implies the ``if'' direction.
	
	To see the ``only if'' direction, let $v_0 = \rec{w[{}:k]}{w}$.
	Then, clearly $\ell_{v_0}=k$ and $x \in \PSuffix(w[{}:k])$ for $x \in v$.
	There must be $v_0,\dots,v_j$ such that $(v_{i-1},v_{i}) \in F$ for $i=1,\dots,j$ and $v_j=v$.
\end{proof}
Therefore, all the p-occurrence positions of the pattern can be found as $\ell_u$ by traversing the descendants $u$ in the reverse suffix link tree $(V,\overline{F})$.
The matching procedure is summarized in Algorithm~\ref{alg:ppdawgmatching}.

The following technical lemma can be used to show that the number of descendants of a node $v$ is linearly bounded by $\msf{RPos}_w(v)$.
\begin{lemma}\label{lem:partition_tree}
	Suppose that each node $u$ of a rooted tree is assigned a nonempty finite set $X_u$ so that 
	\begin{itemize}
		\item if\/ $u$ is a child of\/ $v$, then $X_u \subsetneq X_v$,
		\item if\/ $u$ and $v$ are siblings, then $X_u \cap X_v = \emptyset$.
	\end{itemize}
	Then, $n \le 2|X_r|-1$ where $n$ is the number of nodes of the tree and $r$ is the root.
	Moreover, $n = 2|X_r|-1$ if and only if every leaf $u$ has a singleton set $X_u$ and every inner node $v$ has exactly two children $v_1$ and $v_2$ such that $X_{v_1}$ and $X_{v_2}$ partition $X_v$.
\end{lemma}
\begin{proof}
	We show the lemma by induction on the number of nodes of the tree.
	Suppose $r$ has children $u_1,\dots,u_k$ and let $n_i$ be the number of nodes of the subtree rooted by $u_i$.
	If $k = 0$, since $X_r$ is nonempty, the lemma holds.
	If $k = 1$, by $|X_{u_1}| < |X_r|$, $n = n_1+1$ and the induction hypothesis $n_1 \le 2|X_{u_1}|-1$, we have $n < 2|X_r|-1$.
	If $k \ge 2$, $n = 1+\sum_{i=1}^k n_i \le 1+\sum_{i=1}^k (2|X_{u_i}|-1) \le 2|X_r|-k+1 \le 2|X_r|-1$.
	The equality signs hold only when $k=2$, $|X_r|=|X_{u_1}|+|X_{u_2}|$, and the subtrees rooted by $u_i$ satisfy the stated condition.
\end{proof}
If $(u_1,v),(u_2,v) \in F$, then $|\shortest{u_1}|=|\shortest{u_2}|=|\longest{v}|+1$.
By Lemma~\ref{lem:basic}, $\msf{RPos}_w(u_1) \cap \msf{RPos}_w(u_2) = \emptyset$, unless $u_1=u_2$.
This means that the reversed suffix link tree satisfies Lemma~\ref{lem:partition_tree} for $X_u=\msf{RPos}_w(u)$.
Hence, the number of descendants of $u$ in the reversed suffix link tree is at most $2|\msf{RPos}_w(u)|-1$.
We obtain the following theorem.
\begin{theorem}\label{thm:ppdawg_matching}
  Using $\PPDAWG(T)$ enhanced with the suffix links,
  we can find all substrings of\/ $T$ that p-match a given pattern $P$
  in $O(|P| \log(|\Pi|+|\Sigma|) + \occ)$ time,
  where $\occ$ is the number of occurrences to report.
\end{theorem}

The following theorem contrasts Proposition~\ref{prop:psauto}.
\begin{theorem}\label{thm:ppdawg_node_size}
	If\/ $n=|T| \ge 2$, $\PPDAWG(T)$ has at most $2n-1$ nodes.
	The bound is tight.
\end{theorem}
\begin{proof}
	Let $aw = \lrangle{T}$ with $a \in \Sigma \cup \{\inft\}$ and $w \in (\Sigma \cup \mcal{N})^+$.
	In the reversed suffix link tree $(V,\overline{F})$, $\rec{a}{aw}=\{a\}$ is a child of the root $\rec{\varepsilon}{aw}=\{\varepsilon\}$.
	Suppose $\{\varepsilon\}$ has children $u_1,\dots,u_j$ in addition to $\{a\}$ and $\{a\}$ has children $u_{j+1},\dots,u_k$ in $(V,\overline{F})$, where $\msf{RPos}_w(u_1),\dots,\msf{RPos}_w(u_k)$ are pairwise disjoint and $k \ge 1$ by $n \ge 2$.
	Since only $\varepsilon$ and $a$ can end at the position $1$, $\msf{RPos}_w(u_i) \subseteq \{2,\dots,n\}$.
	Let us partition $V$ into $\{\{\varepsilon\}, \{a\}\}, V_1,\dots,V_k$ where each $V_i$ consists of the nodes of the subtree rooted by $u_i$. 
	By Lemma~\ref{lem:partition_tree}, $|V_i| \le 2|\msf{RPos}_w(u_i)|-1$ and therefore
	\[
		|V| = 2 + \sum_{i=1}^k |V_i| \le 2 + 2 \sum_{i=1}^k |\msf{RPos}_w(u_i)| - k \le 2 + 2 |w| - 1 = 2n-1\,.
	\]
	The tightness is witnessed by a static string $\mtt{ab}^{n-1}$, presented by Blumer et al.~\cite{DAWG}.
\end{proof}
Blumer et al.~\cite{DAWG} have shown in addition that the DAWG for a static string of length $n$ has at most $3n-4$ edges.
We will later show in Corollary~\ref{cor:ppdawg_edge_size} a linear bound on the number of edges of pseudo-PDAWGs.

Our proposed pseudo-PDAWGs are compact enough and support efficient parameterized pattern matching.
In the next section, we will present a modification of pseudo-PDAWGs as our main proposal indexing structure for parameterized strings.

\section{Parameterized directed acyclic word graphs}\label{sec:pdawg_def}
In this section, we present a new indexing structure for parameterized strings, which we call \emph{parameterized directed acyclic word graphs} (\emph{PDAWGs}).
A PDAWG is obtained from a pseudo-PDAWG by suppressing some $\inft$-edges and forgetting the assigned intervals of all $\inft$-edges.
As compensation, we will make use of suffix links for matching.
When two nodes $x_1$ and $x_2$ in $\PSTrie(T)$ are merged into $u=\rec{x_1}{w}=\rec{x_2}{w}$ in $\PPDAWG(T)$, the node $u$ keeps all the outgoing edges of the original nodes $x_1$ and $x_2$.
In $\PDAWG(T)$, we keep only the outgoing edges of $\longest{u}$.
Recall that if $(x_1,a,y_1)$ is an edge of $\PSTrie(T)$ for some $a \in \Sigma \cup \mcal{N} \setminus\{\inft\}$, then $x_2$ also has an $a$-edge $(x_2,a,y_2)$ such that $y_1 \equiv_w^\mrm{R} y_2$.
Therefore, the difference of $\PPDAWG(T)$ and $\PDAWG(T)$ is only in $\inft$-edges.
In this section, we fix a text $T$ and its pv-encoding $w= \lrangle{T}$.
\begin{defn}[Parameterized directed acyclic word graphs]
The \emph{parameterized directed acyclic word graph} (\emph{PDAWG}) $\PDAWG(T)$ of $T$ is a triple $(V,E,F)$ where 
\begin{align*}
	V &= \{\, [x]^\mrm{R}_w \mid  x \in \PFactor(w) \,\}\,,
\\
	E &= \{\,([x]^\mrm{R}_w, a, [xa]^\mrm{R}_w) \in V \times (\Sigma \cup \calN) \times V \mid x = \longest{\rec{x}{w}}\text{ and } xa \in \PFactor(w)   \,\}\,,
\\
	F &= \{\, (u,v) \in (V \setminus \{\{\varepsilon\}\}) \times V \mid v = \rec{\lrangle{x[2:{}]}}{w} \text{ for } x = \shortest{u}\,\}
\,.\end{align*}
Elements of $E$ are called edges and those of $F$ are suffix links.
\end{defn}
The sets $V$ and $F$ remain unchanged from pseudo-PDAWGs.
Note that $(V,E)$ is a directed acyclic graph and $(V,\overline{F})$ is a tree.
Since each $u \in V \setminus \{ [\varepsilon]^\mrm{R}_w \}$ has unique $v$ such that $(u,v) \in F$, we often write $F(u)$ for $v$ regarding $F$ as a function.

\subsection{Parameterized matching based on PDAWGs}
Figure~\ref{fig:pSAoto_pDAWG}(e) shows an example PDAWG.
One may wonder how to find a p-occurrence of $\lrangle{\mtt{axa}}=\mtt{a}\inft\mtt{a}\in \PFactor(\mtt{xaxay})$ using the PDAWG.
Our transition function (Algorithm~\ref{alg:pdawgtransition}) is based on Lemma~\ref{lem:pdawgmatching}, which indicates the node we should visit when the succeeding character is $a \in \Sigma \cup \mcal{N}$ after reading $x$.
The following lemma prepares for Lemma~\ref{lem:pdawgmatching}.
\begin{lemma}\label{lem:pdawgmatch}
For $x \inft \in \PFactor(w)$, let $y = \longest{[x]^\mrm{R}_w}$. Then,
\[
	\msf{RPos}_w(x\inft) = \bigcup \{\, \msf{RPos}_w(y j) \mid j \in \calN \text{ and } j > |x|\,\} \,.
\]
\end{lemma}
\begin{proof}
	To show $\msf{RPos}_w(x\inft) \subseteq \bigcup_{j > |x|} \msf{RPos}_w(y j)$, suppose $i \in \msf{RPos}_w(x\inft)$.
	This implies $i-1 \in \msf{RPos}_w(x) =  \msf{RPos}_w(y)$, $w[i] \in \mcal{N}$, and $w[i] > |x|$.
	Thus, $\lrangle{w[i-|y|:i-1]}=y$ and $i \in \msf{RPos}_w(yj)$ for $j= \Zinf{w[i]}{|y|} \in \{w[i],\inft\}$.
	Hence, we have $j > |x|$.
	
	Conversely suppose $i \in \msf{RPos}_w(yj)$ for some $j > |x|$.
	Then, $i-1 \in \msf{RPos}_w(y)=\msf{RPos}_w(x)$ and $j = \Zinf{w[i]}{|y|} \in \{w[i],\inft\}$.
	On the other hand, $j >|x|$ implies $\Zinf{w[i]}{|x|} = \inft$.
	Thus, $\lrangle{w[i-|x|:i]}=x \inft$, i.e., $i \in \msf{RPos}_w(x\inft)$.
	This proves $\msf{RPos}_w(x\inft) \supseteq \bigcup_{j > |x|} \msf{RPos}_w(y j)$.
\end{proof}

\begin{lemma}\label{lem:pdawgmatching}
Suppose $x \in \PFactor(w)$ and $a \in \Sigma \cup \calN$.
Then, for $y = \longest{[x]^\mrm{R}_w}$,
\[
	[xa]^\mrm{R}_w = \begin{cases}
	[y a]^\mrm{R}_w	& \text{if $a \neq \inft$ or $Z = \emptyset$,}
\\	[y k]^\mrm{R}_w	& \text{if $a = \inft$ and $|Z| = 1$,}
\\	F([y k]^\mrm{R}_w) & \text{if $a = \inft$ and $|Z| \ge 2$,}
	\end{cases}
\]
where 
$
 Z = \{\, j \in \calN \mid y j \in \PFactor(w) \text{ and } j > |x| \,\}
$ and 
$
	k = \min Z
$.
\end{lemma}
\begin{proof}
	If $a \neq \inft$, the lemma immediately follows from Lemma~\ref{lem:ppdawg1}.
	We suppose $a=\inft$.
	If $|Z| \le 1$, we obtain the lemma by Lemma~\ref{lem:pdawgmatch}.
	
	Suppose $|Z| \ge 2$.
	In this case, $|x| < k < \inft$.
	By Lemma~\ref{lem:pdawgmatch}, we see that $\msf{RPos}_w(y k) \subsetneq \msf{RPos}_w(x \inft)$.
	Recall that in general $\msf{RPos}_w(u) \subseteq \msf{RPos}_w(v)$ if and only if $v$ is reachable from $u$ by following a certain number (including zero) of suffix links in $(V,F)$.
	Hence, one can reach $[x\inft]_w^\mrm{R}$ from $[yk]_w^\mrm{R}$ by following at least one suffix link.
	To show that there is no other node between $[x\inft]_w^\mrm{R} $ and $[yk]_w^\mrm{R}$, it suffices to show that
	for any $z \in \PSuffix(yk)$ such that $|x\inft| < z < |yk|$ (and thus $\msf{RPos}_w(yk) \subseteq \msf{RPos}_w(z) \subseteq \msf{RPos}_w(x\inft)$), either $\msf{RPos}_w(z) = \msf{RPos}_w(x \inft)$ or $\msf{RPos}_w(z) = \msf{RPos}_w(yk)$.
	Here, $z$ must be of the form $z = z' \Zinf{k}{|z'|}$.
	The assumption implies $\longest{[x]_w^\mrm{R}}= \longest{[z']_w^\mrm{R}}=\longest{[y]_w^\mrm{R}} $.
	Note that $yk \in \PFactor(w)$ and $k < \inft$ implies $|y| \ge k$.

	Suppose $|x\inft| < |z| \le k $, i.e., $z = z' \inft$.
	By Lemma~\ref{lem:pdawgmatch}, $|z'| < k$ implies $\msf{RPos}_w(z' \inft) = \msf{RPos}_w(x \inft) = \bigcup_{j \geq k} \msf{RPos}_w(yj) $ by the choice of $k$.
	
	Suppose otherwise, $k < |z| < |yk|$, i.e., $z = z'k$.
	Lemma~\ref{lem:ppdawg1} implies $\msf{RPos}_w(z' k) = \msf{RPos}_w(y k)$.
\end{proof}

\begin{algorithm2e}[t]
	\caption{Function $\msf{trans}(u,i,a)$\label{alg:pdawgtransition}}
	\lIf{$a \ne \inft$}{%
		\textbf{return} $\msf{child}(u,a)$%
	}\Else{%
		Let $Z \leftarrow \{\, j \in \calN \cap \msf{Children}(u) \mid j > i\,\}$\;
	\lIf{%
		$Z = \emptyset$%
	}{\textbf{return} $\msf{Null}$}
	\lElseIf{%
		$Z = \{b\}$%
	}{\textbf{return} $\msf{child}(u,b)$}
	\lElse{%
		\textbf{return} ${F}(\msf{child}(u,b))$ for $b = \min Z$%
	}}
\end{algorithm2e}
\begin{algorithm2e}[t]
	\caption{Parameterized pattern matching algorithm based on $\PDAWG(T)$\label{alg:pdawgmatching}}
\let\oldnl\nl
\newcommand{\nonl}{\renewcommand{\nl}{\let\nl\oldnl}}
	\SetVlineSkip{0.5mm}
	$p \leftarrow \lrangle{P}$\;
	Let $u$ be the source node of $\PDAWG(T)$\;
	\For{$i = 1$ \textup{\textbf{to}} $|P|$}{%
		Let $u \leftarrow \msf{trans}(u,i-1,p[i])$\;
		\lIf{$u = \msf{Null}$}{\textbf{return} $\mathsf{False}$}
	}
	Traverse the reversed suffix link tree and \\ \nonl \quad \textbf{output} $\ell_v$ for all descendants $v$ of $u$\;
\end{algorithm2e}

The function $\msf{trans}$ of Algorithm~\ref{alg:pdawgtransition} is a straightforward realization of Lemma~\ref{lem:pdawgmatching}, where $\msf{Children}(u)$ denotes the set of labels of the outgoing edges of $u$
and $\msf{child}(u,a)$ is the node that the $a$-edge of $u$ points at.
If $u$ has no $a$-edge, $\msf{child}(u,a)=\msf{Null}$. 
In other words, $\msf{Children}(u) = \{\, a \in \calN \cup \Sigma \mid (u,a,v) \in E$ for some $v \in V \,\}$ and $\msf{child}(u,a) = v \in V$ iff $(u,a,v) \in E$.
The algorithm takes a node $u \in V$, a natural number $i \in \mbb{N}$, and a character $a \in \Sigma \cup \calN$,
 and returns the node where we should go by reading $a$ from $u$ assuming that we have read $i$ characters so far.
By Lemma~\ref{lem:pdawgmatching}, $\msf{trans}([x]^\mrm{R}_w,|x|,a) = [xa]^\mrm{R}_w$ for every $xa \in \PFactor(w)$.
On the other hand, suppose $x \in \PFactor(T)$ and $xa \notin \PFactor(T)$.
If $a \neq \inft$, $ya \notin \PFactor(T)$ for $y=\rlong{x}{w}$.
If $a=\inft$, for any $i \in \msf{RPos}_w(x) = \msf{RPos}_w(y)$, either $w[i+1] \in \Sigma$ or $w[i+1] \le |x|$.
Thus, the node $[x]^\mrm{R}_w = [y]^\mrm{R}_w$ has no edge labeled with a character in $\calN$ greater than $|x|$.
The algorithm returns \textsf{False}. 
Using $\msf{trans}$, Algorithm~\ref{alg:pdawgmatching} performs p-matching, where $\ell_v = \min \msf{RPos}(v)$.

\begin{theorem}\label{thm:matching}
  Using $\PDAWG(T)$, we can find all p-occurrences of $P$ in $T$ in $O(|P| \log(|\Pi|+|\Sigma|) + \occ)$ time,
  where $\occ = |\msf{RPos}_{\lrangle{T}}(\lrangle{P})|$ is the number of occurrences to report.
\end{theorem}

\subsection{Size of PDAWGs}
Blumer et al.~\cite{DAWG} have shown that the DAWG for a static string of length $n$ has at most $2n-1$ nodes and $3n-4$ edges.
We show that PDAWGs have the same size bound.
\begin{theorem}\label{thm:pdawg_size}
	$\msf{PDAWG}(T)$ has at most $2n-1$ nodes and $3n-4$ edges when $n=|T| \ge 3$.
	Those bounds are tight.
\end{theorem}
\begin{proof}
	Concerning the number of nodes, Theorem~\ref{thm:ppdawg_node_size} holds for PDAWGs, since the node sets of PDAWGs and pseudo-PDAWGs are identical.
	
	On the number of edges, we first give a weaker upper bound $3n-3$, just like Blumer et al.~\cite{DAWG} have done.
	Let $\msf{PDAWG}(T) = G = (V,E,F)$, $\msf{PSTrie}(T) = H = (U,D)$, $V' \subsetneq V$ the set of non-sink nodes of $G$, $U' \subsetneq U$ the set of internal nodes of $H$, and $d_G(v)$ denote the out-degree of node $v$ in $G$.
	We have $|E| = \sum_{v \in V'} d_G(v) = \sum_{v \in V'} d_{H}(\longest{v})$, since $d_{G}(v)=d_{H}(\longest{v})$ for all $v \in V$.
	Since $H$ has at most $n$ leaves, 
	\begin{align*}
	n \ge |U \setminus U'| = 1+\sum_{u \in U'}(d_H(u)-1) \ge 1+\sum_{v \in V'}(d_H(\longest{v})-1) = 1+|E|-|V'| \,, 
	\end{align*}
	which implies $|E| \le n+|V'|-1 \le 3n-3$.

	This upper bound $3n-3$ could be achieved only when $|V| = 2n-1$.
	We will show that if $|V|=2n-1$, then the skeleton (stripping off edge labels) of $\msf{PDAWG}(T)$ is isomorphic to that of $\msf{PDAWG}(\mtt{a}\mtt{b}^{n-1})$ for $\mtt{a},\mtt{b} \in \Sigma$ with $\mtt{a} \neq \mtt{b}$, where the source is the only branching node from which two paths of length $n$ and $n-1$ reach the sink.
	
	Let $aw=\lrangle{T}$ with $a \in \Sigma \cup \{\inft\}$ and $\overline{F}(u)=\{\,v \mid (v,u) \in F\,\}$ for $u \in V$, i.e., $\overline{F}(u)$ is the set of children of $u$ in the reversed suffix link tree.
	According to the proof of Theorem~\ref{thm:ppdawg_node_size}, $|V|=2n-1$ can be achieved only when $\overline{F}(\{\varepsilon\}) \cup \overline{F}(\{a\}) \setminus \{\{a\}\} = \{u\}$ for some $u \in V$ such that $\msf{RPos}_{aw}(u)=\{2,\dots,n\}$.
	Moreover, by Lemma~\ref{lem:partition_tree}, it must hold that $\overline{F}(u)=\{u_1,u_2\}$ and that $\msf{RPos}_{aw}(u_1)$ and $\msf{RPos}_{aw}(u_2)$ partition $\msf{RPos}_{aw}(u)=\{2,\dots,n\}$.
	Suppose $\overline{F}(\{\varepsilon\})=\{\{a\}\}$ and $\overline{F}(\{a\})=\{u\}$.
	In this case, by $\msf{RPos}_{aw}(u) \subsetneq \msf{RPos}_{aw}(a) \subsetneq \msf{RPos}_{aw}(\varepsilon)$,
	it must hold $\msf{RPos}_{aw}(a) = \{1,\dots,n\}$.
	Since there are at most $k+1$ pv-strings $x$ such that $k \in \msf{RPos}_{\lrangle{S}}(x)$ for any p-string $S$ in general,
	there can be at most three nodes $v$ in $V$ such that $2 \in \msf{RPos}_{aw}(v)$, which are actually $\{\varepsilon\}$, $\{a\}$ and $u$.
	Therefore, $\msf{RPos}_{aw}(u_1)$ and $\msf{RPos}_{aw}(u_2)$ cannot partition $\msf{RPos}_{aw}(u)$.
	Consequently, it must hold  $\overline{F}(\{\varepsilon\})=\{\{a\},u\}$ and $\overline{F}(\{a\})=\emptyset$.
	Since $\msf{RPos}_{aw}(\{a\}) \cap \msf{RPos}_{aw}(u) = \emptyset$, we must have $\msf{RPos}_{aw}(\{a\}) =\{1\}$.
	By $\rec{w[k:k]}{aw} \in \overline{F}(\{\varepsilon\})$ for all $k$, we must have
	either $a \in \Sigma$ and $w \in \mcal{N}^{n-1}$ or $w = b^{n-1}$ for some $b \in \Sigma^{n-1}$.
	In the latter case, it is easy to see that $\PDAWG(T)$ is isomorphic to $\PDAWG(\mtt{a}\mtt{b}^{n-1})$.
		
	So, hereafter we assume $a \in \Sigma$ and $w \in \mcal{N}^{n-1}$.
	Then, $\msf{RPos}_w((aw)[:k]) = \{k\}$ and $[(aw)[:k]]^\mrm{R}_w = \{(aw)[:k]\}$ for all $k \in \{1,\dots, n\}$.
	We show by induction on $k$ that $\msf{RPos}_w(\lrangle{w[i:i+k-1]})=\{k+1,\dots,n\}$ for $1 \le i \le n-k$ for $1 \le k \le n-1$, i.e., $\PFactor(w)$ has just one pv-string of length $k$.
	This is true for $k=1$, since $\lrangle{w[i]}=\inft$ for $1 \le i \le n-1$.
	This is also trivially true for $k=n-1$.
	Suppose the claim holds up to $k \le n-3$.
	According to the proof of Lemma~\ref{lem:partition_tree}, to achieve the tight upper bound $|V|=2n-1$, it must hold $\overline{F}([w[:k]]^\mrm{R}_w)=\{u_1,u_2\}$ such that $\msf{RPos}(u_1)$ and $\msf{RPos}(u_2)$ partition $\msf{RPos}_w(w[:k]) = \{k+1,\dots,n\}$.
	Since $F([a (w[:k])]^\mrm{R}_w) = [w[:k]]^\mrm{R}_w$ and $\msf{RPos}_w(a (w[:k])) = \{k+1\}$, it must hold $\msf{RPos}(u_2)=\{k+2,\dots,n\}$ for $\overline{F}([w[:k]]^\mrm{R}_w)=\{[a (w[:k])]^\mrm{R}_w,u_2\}$.
	Take an arbitrary element $x \in u_2$.
	The fact $k+2,n \in \msf{RPos}_w(x)$ means $x = \lrangle{w[i:k+2]}$ for some $i \ge 1$.
	By induction hypothesis, any element of $\msf{PFactor}(w)$ of length $j \le k$ is $\lrangle{w[:j]}$.
	Therefore, $x$ must have length $k+1$, i.e., $x=w[:k+1]$ and thus $u_2 = \{w[:k+1]\}$.
	$\msf{RPos}_{aw}(u_2)=\{k+1,\dots,n\}$ means $w[:k+1]=\lrangle{w[i:k+i-1]}$ for $1 \le i < n-k$.
	Summarizing above, all nodes of $V$ are $[w[:k]]^\mrm{R}_w$ and $[a (w[:k])]^\mrm{R}_w$ for $0 \le k \le n-2$ and the sink node $\{w[:n-1],a (w[:n-1])\}$.
	This PDAWG is isomorphic to $\PDAWG(\mtt{a}\mtt{b}^{n-1})$, which has $2n-1$ edges.

	One can achieve $|E|=3n-4$ by the string $\mtt{a}\mtt{b}^{n-2}\mtt{c}$, given by Blumer et al.~\cite{DAWG}.
\end{proof}
\begin{corollary}\label{cor:ppdawg_edge_size}
	$\msf{pPDAWG}(T)$ has at most $5n-7$ edges when $n=|T| \ge 3$.
\end{corollary}
\begin{proof}
	By Theorem~\ref{thm:pdawg_size}, it suffices to evaluate the number of $\inft$-edges of $\PPDAWG(T)$ which do not appear in $\PDAWG(T)$.
	We show that each node of $\PPDAWG(T)$ has at most one such incoming $\inft$-edge.
	Suppose a (non-source) node $v$ has two incoming $\inft$-edges $(u_1,\inft,v)$ and $(u_2,\inft,v)$ in $\PPDAWG(T)$.
	This means that there are $x_1 \in u_1$ and $x_2 \in u_2$ such that $x_1 \inft, x_2 \inft \in v$.
	We assume without loss of generality that $|x_1| < |x_2|$, which implies $|x_1| \le |\longest{u_1}|<|\shortest{u_2}| \le |x_2|$.
	Then, $\longest{u_1}\inft \in v$ by Lemma~\ref{lem:basic}, which means that $\PDAWG(T)$ has the edge $(u_1,\inft,v)$.
	That is, only $(u_2,\inft,v)$ may be suppressed in $\PDAWG(T)$.
	Therefore, the number of edges of $\PPDAWG(T)$ is at most $|E|+|V|-1 \le 3n-4+2n-2-1=5n-7$ for $\PDAWG(T)=(V,E,F)$.
\end{proof}
It is an open problem to give the tight upper bound on the number of edges of pseudo-PDAWGs.

\section{Duality of PDAWGs and p-suffix trees}\label{sec:duality}

This section establishes the duality between parameterized suffix trees~\cite{Baker93} and PDAWGs.
  This will give us the following two merits:
  the \emph{first} bidirectional index for
  parametrized pattern matching (Section~\ref{sec:Weiner_links}),
  and efficient offline construction of PDAWGs (Section~\ref{sec:pdawg_pst}).

\subsection{Parameterized suffix trees and Weiner links}
\label{sec:Weiner_links}

In this subsection, we first recall the basic properties of p-suffix trees (Section~\ref{sec:basics_PST}),
and then the suffix links of p-suffix trees (Section~\ref{sec:suffix_link}).
Then, we introduce our Weiner links of p-suffix trees (Section~\ref{sec:weiner_PST}),
and show that the Weiner links are equal to the PDAWG edges (Section~\ref{sec:duality_bidirection}).
This immediately leads us to bidirectional indexing structure for p-matching.

\subsubsection{Basics of p-suffix trees}\label{sec:basics_PST}

\begin{figure}[t]
  \centering
  	\begin{minipage}[t]{0.49\hsize}
		\centering
		\includegraphics[scale=0.22]{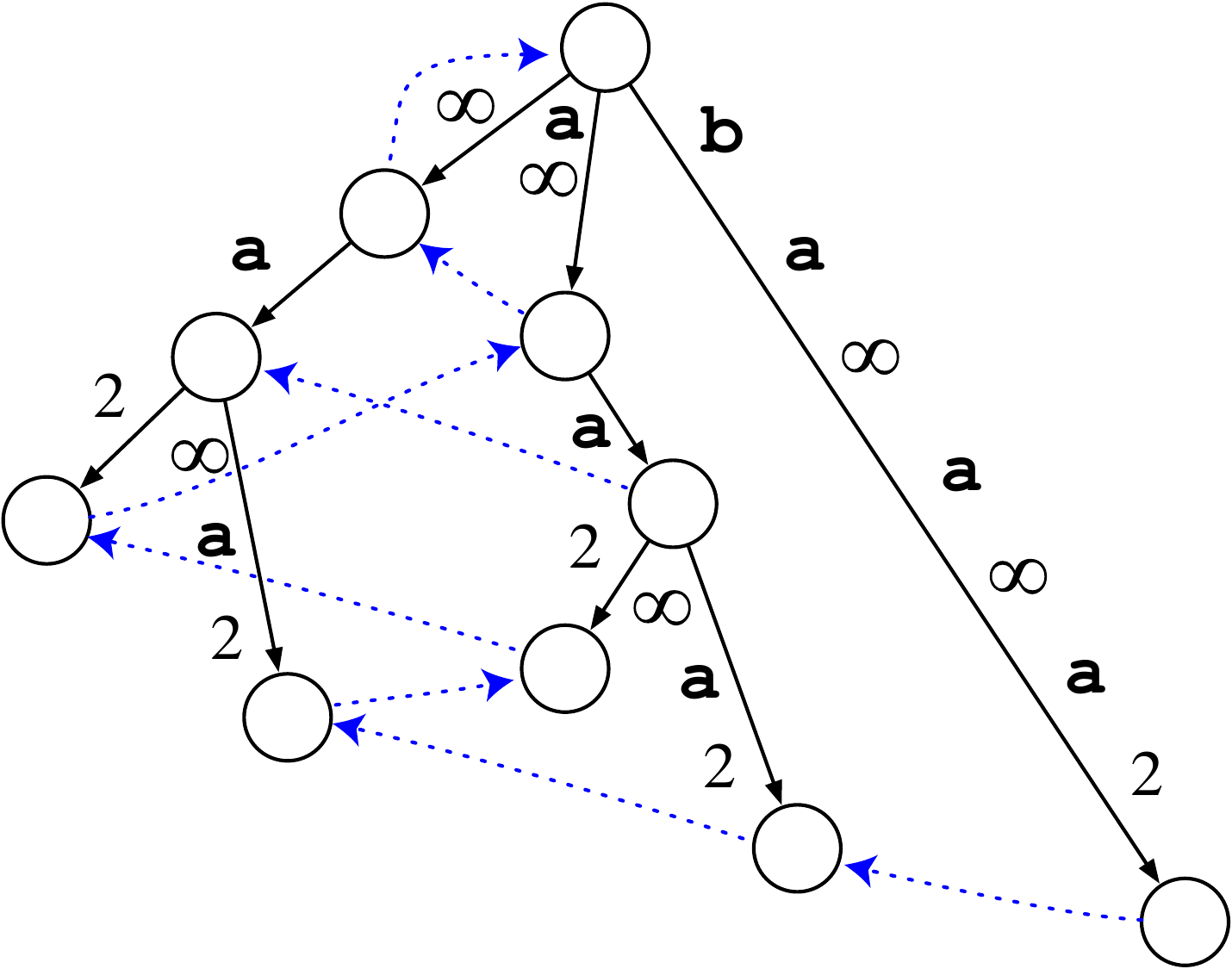}\\
		\ \ \ \scriptsize{(a)}
	\end{minipage}
	\begin{minipage}[t]{0.49\hsize}
		\centering
		\includegraphics[scale=0.22]{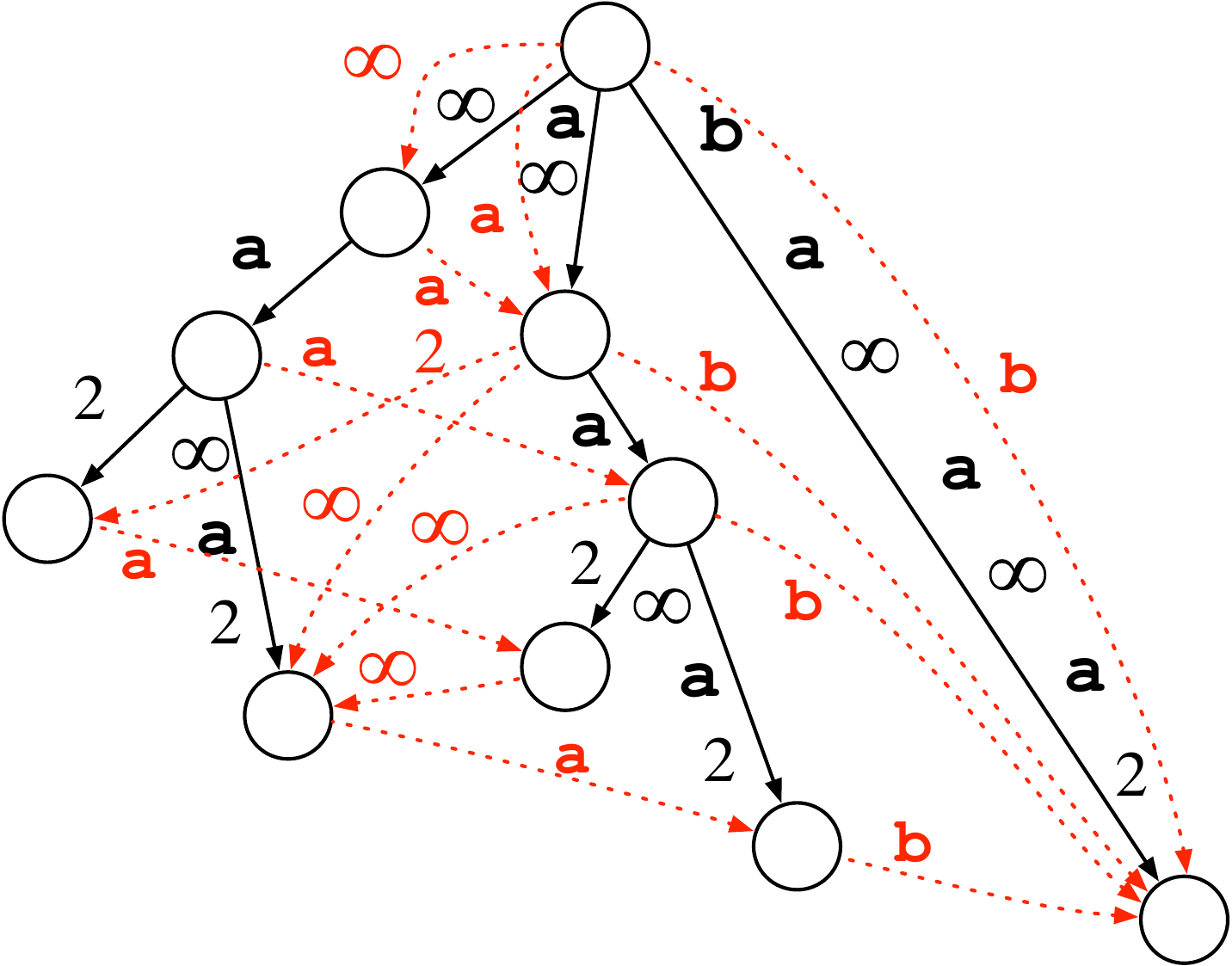}\\
		\ \ \ \scriptsize{(b)} \vspace*{1pc}
	\end{minipage}
	\begin{minipage}[t]{0.49\hsize}
		\centering
		\includegraphics[scale=0.22]{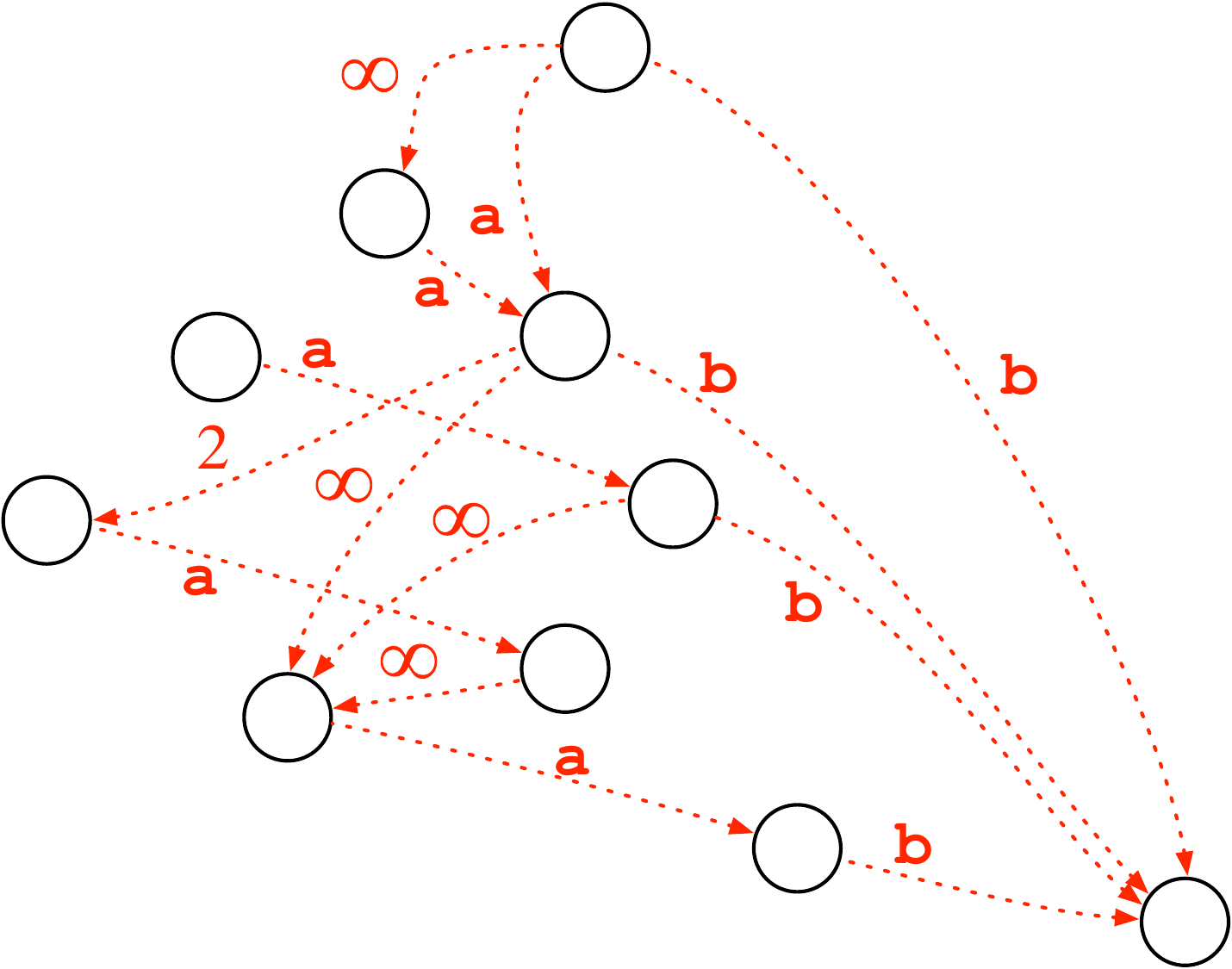}\\
		\ \ \ \scriptsize{(c)}	
	\end{minipage}
	\begin{minipage}[t]{0.49\hsize}
		\centering
		\includegraphics[scale=0.22]{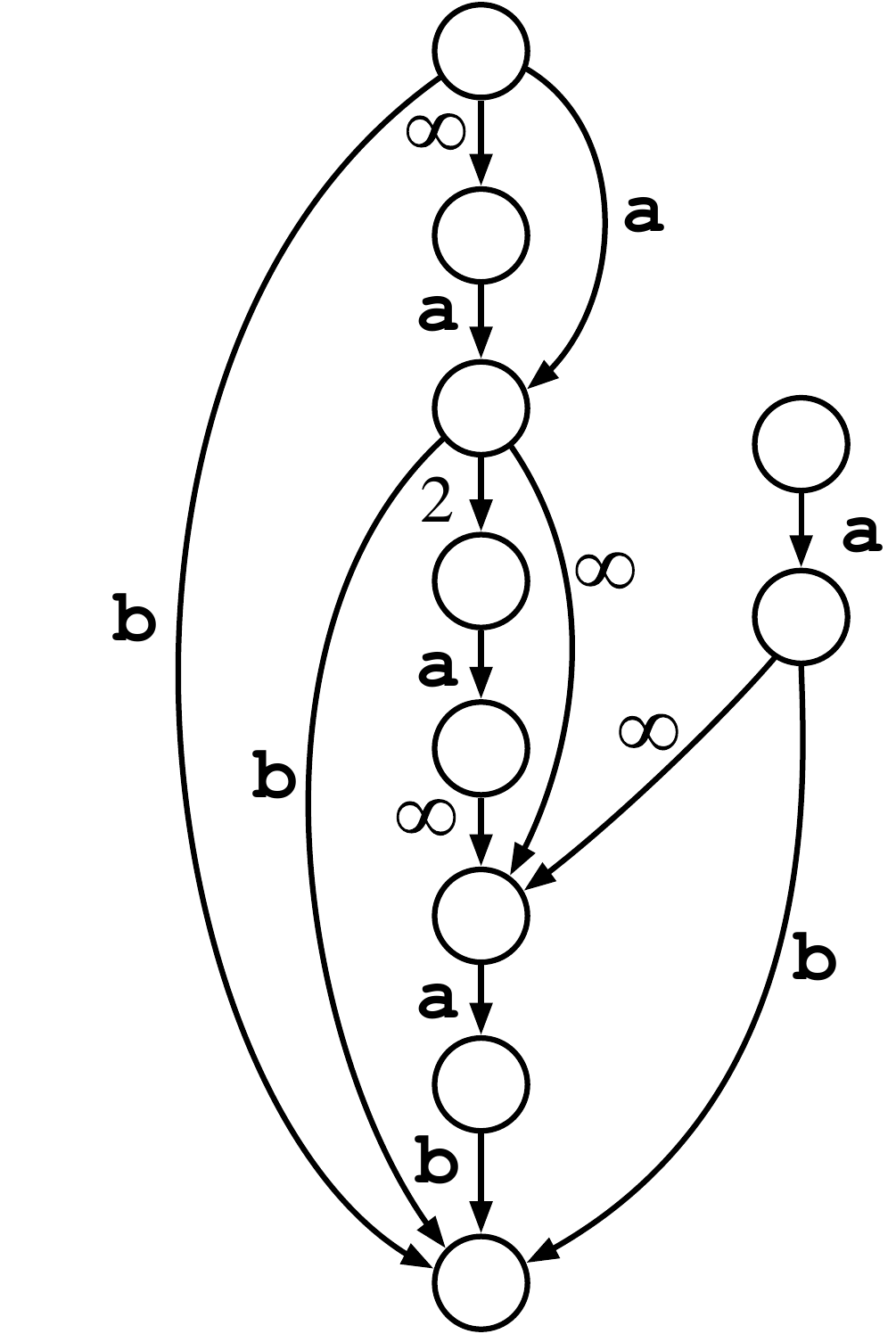}\\
		\ \ \ \scriptsize{(d)}	
	\end{minipage}
	\caption{\label{fig:pSTree_pDAWG}
          Consider string $T = \mathtt{yayaxab}$ over $\Sigma = \{\mathtt{a, b}\}$ and $\Pi = \{\mathtt{x, y}\}$.
         (a) The parameterized suffix tree $\PSTree(\rev{T})$ with $\rev{T} = \mathtt{baxayay}$ augmented with the suffix links (dashed blue arcs).
	 (b) The parameterized suffix tree $\PSTree(\rev{T})$ augmented with the Weiner links (dashed red arcs).
         (c) The DAG consisting of the p-suffix tree nodes and the Weiner-links.    	  (d) The PDAWG $\PDAWG(T)$ with $T = \mathtt{yayaxab}$.
	 Observe that the edge-labeled graphs (c) and (d) are isomorphic.
	}	
\end{figure}

Let $T$ be a p-string and consider its reversal $\rev{T}$.
The \emph{parameterized suffix tree (p-suffix tree)} $\PSTree(\rev{T})$ of $\rev{T}$ is the path-compacted (or Patricia) tree for $\PSuffix(\rev{T})$.
Below let us recall the definition of $\PSTree(\rev{T})$,
which is the edge-labeled tree $(V,E)$ of $\rev{T}$ such that for $w=\lrangle{\rev{T}}$
\begin{align*}
	V &= \{\, \llong{x}{w} \mid x \in \PFactor(\rev{T}) \,\},
\\	E &= \{\, (x,y,xy) \in V \times (\Sigma \cup \calN)^+ \times V  \mid xy = \llong{xa}{w} \in V \text{ for some } a \in \Sigma \cup \mcal{N}\,\}.
\end{align*}
Recall that each edge of $\PSTree(T)$ is labeled by an element of $\Factor(\PSuffix(T)) \setminus \{\varepsilon\}$.
An example for $\PSTree(\rev{T})$ is given in Figure~\ref{fig:pSTree_pDAWG}(a).

\begin{remark}
  Since each node of $\PSTree(\rev{T})$ is defined as the longest member $\llong{x}{w}$
  of the equivalence class $[x]_w^{\mathrm {L}}$,
  $\PSTree(\rev{T})$ may contain an internal node that has only a single child.
  Such an internal node corresponds to a suffix of $\rev{T}$
  that has internal p-matching occurrences in $\rev{T}$.
  For instance, $\mathtt{a\inft}$ of the parameterized suffix tree shown
  in Figure~\ref{fig:pSTree_pDAWG}(a) has only a single child.
  This is because $\mathtt{a\inft}$ has three p-matching occurrences in $\mathtt{baxayay}$
  at positions $2, 4, 6$,
  where the last occurrence corresponds to the suffix $\mathtt{ay}$
  and all of the other internal p-matching occurrences of $\mathtt{a\inft}$
  are immediately followed by $\mathtt{a}$.
  If we use a common convention that $\rev{T}$ terminates with a unique character $\$$,
  all internal nodes of $\PSTree(\rev{T})$ become branching.
\end{remark}

Recall that each node of $\PSTree(\rev{T})$ is a pv-string.
The \emph{string depth} of a node $x$ of $\PSTree(\rev{T})$ is $|x|$.
We store the string depth $|x|$ in each node $x$ of $\PSTree(\rev{T})$.
To represent $\PSTree(\rev{T})$ in linear space, in an actual implementation, the label $y$ of an edge $(x,y,xy)$ is represented by two integers $i$ and $j$ such that $y=\lrangle{\rev{T}[i-|x|:j]}[|x|+1:j-i+|x|+1]$.
In other words, $y$ is the length-$(j-i+1)$ suffix of the pv-encoding of $\langle \rev{T}[i-|x|:j] \rangle = xy$
which labels the path from the root to the node $xy$.

\subsubsection{Suffix links of p-suffix trees} \label{sec:suffix_link}

Let us recall the \emph{suffix links} of the p-suffix tree,
which were first introduced by Baker~\cite{Baker93}.
Let $u$ be a non-root node of $\PSTree(\rev{T})$
and let $S$ be any substring of $\rev{T}$ such that $\lrangle{S} = u$.
The suffix link of $u$, denoted $\ssl(u)$,
is a pointer from $u$ to $v = \lrangle{S[2:|S|]}$.
Notice that $v$ may not be a node of the p-suffix tree.
For instance, see Figure~\ref{fig:pSTree_pDAWG}(a).
Consider node $u = \inft \mathtt{a}$, in which we have $S = \mathtt{ya}$ or $S = \mathtt{xa}$.
In either case $v = \lrangle{S[2:2]} = \lrangle{a} = \mathtt{a}$,
however, there is no node representing $\mathtt{a}$.
In this paper, we define the suffix link of a node $u = \lrangle{S}$ only if $v = \lrangle{S[2:|S|]}$ is a node.

We can characterize the target node $v$ of the suffix link $\ssl(u)$ depending on the first character $u[1]$ of $u$, as follows:
\begin{enumerate}
\item[(i)] If $u[1] \in \Sigma$, then $v = u[2:|u|]$. This is the same as the suffix link of a standard suffix tree for exact matching.

\item[(ii)] If $u[1] = \inft$ and there exists a position $a$ in $u$ such that
   $u[a] = a-1$, then $u[a]$ ``points'' to the first position of $u$ in the prev-encoding. Notice that such a position $a$ is unique and that $a$ is the smallest position in $S$ that is larger than $1$ with $S[1] = S[a]$. Now, the target node $v$ is $u[2:a-1] \cdot \inft \cdot u[a+1:|u|]$, that is obtained by removing $u[1]$ from $u$ and replacing $u[a]$ with $\inft$.

\item[(iii)] If $u[1] = \inft$ and there is no position $a$ in $u$ such that
   $u[a] = a-1$, then $v = u[2:|u|]$.
\end{enumerate}
For examples of suffix links, see Figure~\ref{fig:pSTree_pDAWG}(a).
The suffix link of node $\mathtt{a} \inft \mathtt{a}$ points to node $\inft \mathtt{a}$, which is Case (i).
The suffix link of node $\inft \mathtt{a}2$ points to node $\mathtt{a} \inft$, which is Case (ii).
The suffix link of node $\inft \mathtt{a} \inft \mathtt{a}2$ points to node $\mathtt{a} \inft \mathtt{a}2$, which is Case (iii).

\subsubsection{Weiner links of p-suffix trees}\label{sec:weiner_PST}

We enhance p-suffix trees by introducing \emph{Weiner links},
which are key tools in this whole section.
The \emph{reversals} of the suffix links of Section~\ref{sec:suffix_link}
are explicit Weiner links.
In Case (i), the corresponding explicit Weiner link is labeled with the static character $u[1] \in \Sigma$.
In Case (iii), the corresponding explicit Weiner link is labeled with $u[1] = \inft$.
In Case (ii), we label the corresponding explicit Weiner link with the position $a$.
In a unified manner for all these three cases, we can represent each explicit Weiner link by a triple $(v, a, u)$, where $a \in \Sigma \cup \calN$ and $|u| = |v|+1$.

Now, we are to extend the notion of Weiner links to the case where there is no node $u$ with $|u| = |v|+1$ for a given node $v$ and $a \in \Sigma \cup \calN$. In such a case, we use the shortest (i.e.\ shallowest) possible node as $u$, as follows.
Let $v$ be a node in $\PSTree(\rev{T})$ such that $v = \lrangle{S}$ for some substring $S$ of $\rev{T}$, and $a \in \Sigma \cup \calN$.
Let $\alpha(a, v)$ be the pv-string such that
\begin{equation}
\alpha(a, v) =
\begin{cases}
  av & \mbox{if } a \in \Sigma \cup \{\inft\} \mbox { and } av \in \PFactor(T), \\
  \lrangle{S[a] \cdot S} & \mbox{if } a \in \calN \setminus \{\inft\}  \mbox{ and } \lrangle{S[a] \cdot S} \in \PFactor(T), \\
  \mbox{undefined} & \mbox{otherwise}.
\end{cases}
\label{def:Weiner_label}
\end{equation}
This function $\alpha$ corresponds to the reversals of the suffix links.
When $a \in \Sigma$, it ``prepends'' label $a$ to string (node) $v$.
When $a \in \calN$, it gives the pv-encoding of the p-string which is obtained by prepending the parameter character indicated by $a$ to the p-string $S$ whose pv-encoding is $v$.
For $a = \inft$, the indicated parameter character is a fresh one that occurs nowhere in $S$.  
For $a \in \calN \setminus \{\inft\}$, the parameter character is $S[a]$.
Then, the Weiner link from node $v$ to node $u =\longest{\lec{\alpha(a,v)}{\lrangle{\rev{T}}}}$ is labeled with $a$, which is represented by the triple $(v,a,u)$.
Note that the operator $\longest{\lec{\cdot}{\lrangle{\rev{T}}}}$ brings us
to the shallowest node $u$ from the locus for $\alpha(a,v)$ in $\PSTree(\rev{T})$.
Thus, each Weiner link increases the string depth by at least one,
and hence $|u| \geq |v|+1$ always holds.

The Weiner link $(v,a,u)$ is said to be 
\emph{explicit} if $|u| = v+1$
(or equivalently $u = \alpha(a, v)$),
and \emph{implicit} if $|u| > |v|+1$.
Namely, $u = \alpha(a, v)$ if and only if
$v$ is obtained by simply removing the first character $a$ from $u = av$
(the first case in equation~(\ref{def:Weiner_label})),
or by replacing $u[a+1] = a$ with $\inft$
and removing the first character $\inft$ from $u$ (the second case in equation~(\ref{def:Weiner_label})).
Basically the same arguments hold for implicit Weiner links,
except in that we need to cut off the
suffix of $u$ to adjust the length to $|v|+1$.

For examples of Weiner links of a p-suffix tree,
see Figure~\ref{fig:pSTree_pDAWG}(b).
The node $v = \mathtt{a\inft}$ has
an explicit Weiner link which is labeled ${2}$ and points to the node
$u = \inft \mathtt{a}2$.
This is because by replacing $u[2+1] = {2}$ with $\inft$
and by removing the first character $\inft$ from $u$,
we obtain $v = \mathtt{a\inft}$.
This corresponds to the second case in equation~(\ref{def:Weiner_label}).
For another example, consider the node $u' = \mathtt{\inft a\inft a}2$
which has three in-coming Weiner links all labeled $\inft$.
The Weiner link from the node $v' = \mathtt{a\inft a}2$ is explicit,
while the other two from the node $v''= \mathtt{a\inft a}$ and $v'''= \mathtt{a\inft}$
are implicit.
Note that all these Weiner links correspond to 
the first case in equation~(\ref{def:Weiner_label}),
namely, one can obtain $v'$ by simply removing the first $\inft$ from $u$,
and can obtain $v''$ and $v'''$ by removing the first $\inft$ from $u$
and removing the suffixes of $u$ accordingly.

\subsubsection{Duality between PDAWGs and p-suffix trees and bidirectional searches}\label{sec:duality_bidirection}

Our Weiner links for $\PSTree(\rev{T})$ permit us
to design a Weiner-style~\cite{ST} right-to-left online construction of $\PSTree(\rev{T})$,
which turns out to be equivalent to
our left-to-right online construction of $\PDAWG(T)$
to be presented in Section~\ref{sec:pdawg_construction}.
This observation is based on the following duality
between $\PSTree(\rev{T})$ and $\PDAWG(T)$.

To establish the correspondence between $\PDAWG(T)$ and $\PSTree(\rev{T})$,
we define the \emph{p-reverse} $\pr{x}$ of a pv-string $x$ so that $\pr{x} = \lrangle{\rev{S}}$ iff ${x} = \lrangle{S}$ for any p-string $S \in (\Sigma \cup \Pi)^*$.
 For example, for $T=\mtt{xaxy}$ with $\mtt{a} \in \Sigma$ and $\mtt{x},\mtt{y} \in \Pi$, we have 
$\pr{\lrangle{T}}=\pr{\inft\mtt{a}2\inft}=\inft\inft\mtt{a}2 = \lrangle{\mtt{yxax}}= \lrangle{\rev{T}}$.
 
For technical convenience, we rename the nodes $\rec{x}{\lrangle{T}}$ of $\PDAWG({T})$ to be $\longest{\rec{x}{\lrangle{T}}}$ in this section.
Moreover, we call an edge $(x,a,y)$ of $\PDAWG({T})$ \emph{primary} if $y=xa$, and \emph{secondary} otherwise.
\begin{theorem}\label{lem:duality}
The following correspondence between $\msf{PDAWG}(T) = (V_\msf{D}, E_\msf{D})$ and $\msf{PSTree}(\rev{T}) = (V_\msf{T},E_\msf{T})$ holds.
\begin{enumerate}
	\item[(1)] $\msf{PDAWG}(T) $ has a node $x \in V_\msf{D}$ iff\/ $\msf{PSTree}(\rev{T})$ has a node $\pr{x} \in V_\msf{T}$.
	\item[(2)] $\msf{PDAWG}(T) $ has a primary edge $(x,a,y) \in E_\msf{D}$ iff\/ $\msf{PSTree}(\rev{T})$ has an explicit Weiner link $(\pr{x},a,\pr{y})$.
	\item[(3)] $\msf{PDAWG}(T) $ has a secondary edge $(x,a,y) \in E_\msf{D}$ iff\/ $\msf{PSTree}(\rev{T})$ has an implicit Weiner link $(\pr{x},a,\pr{y})$.
	\item[(4)] $\msf{PDAWG}(T) $ has a suffix link from ${xy}$ to ${y}$ iff\/ $\msf{PSTree}(\rev{T})$ has an edge $(\pr{y}, \pr{x}, \pr{xy}) \in E_\msf{T}$.
\end{enumerate}
\end{theorem}
\begin{proof}
	To make the arguments simpler,
	we assume for now that $T$ begins with a unique character $\$$
	that does not occur elsewhere in $T$.
	The case without $\$$ can be shown similarly.
	
	\begin{enumerate}
          \item[(1)] By the symmetry $\msf{RPos}_{\lrangle{T}}(x) = \{\, n+1-k \mid k \in \msf{LPos}_{\lrangle{\rev{T}}}(\pr{x})\,\}$,
	we have $x=\longest{\rec{x}{\lrangle{T}}}$ if and only if $\pr{x}=\longest{[\pr{x}]_{\lrangle{\rev{T}}}^{\mrm{L}}}$.
	To see why this holds more intuitively,
        let $\parent(\pr{x})$ be the parent of $\pr{x}$ in $\PSTree(\rev{T})$,
	and let $\ell$ be the edge label from $\parent(\pr{x})$ to $\pr{x}$.
	Then, for any locus on this edge representing
	$\pr{z_i} = \parent(\pr{x}) \cdot \ell[1:i]$,
	with $1 \leq i \leq |\ell|$, there are the same leaves below it.
	Since each leaf of $\PSTree(\rev{T})$ corresponds to a distinct position in $\lrangle{\rev{T}}$,
	every $\pr{z_i}$ has the same set of beginning positions in $\lrangle{\rev{T}}$
	(note that $\pr{z_\ell} = \pr{x}$).
	By symmetry, this in turn means that 
	$z_i$ has the same set of ending positions in $\pr{\lrangle{\rev{T}}} = \lrangle{T}$,
	i.e. $\{z_i \mid 1 \leq i \leq |\ell|\} = \rec{x}{\lrangle{T}}$.
	The other way (from PDAWG nodes to p-suffix tree nodes)
	can be shown analogously.

	\item[(2)] Because $(\pr{x},a,\pr{y})$ is an explicit Weiner link,
	$\alpha(a, \pr{x}) = \pr{y}$.
	By the definition of operator $\pr{\cdot}$, we obtain $xa = y$.
	Hence there is a primary edge from node $x$ to $y$ labeled $a$
	in $\PDAWG(T)$.
	The other way (from PDAWG primary edges to p-suffix tree explicit Weiner links)
	can be shown analogously.
        
	\item[(3)] Similar to (2).
	\item[(4)] Immediately follows from the proof for (1).
      \end{enumerate}
\end{proof}

Properties (2) and (3) imply that
we can use the Weiner links of $\PSTree(\rev{T})$
for parameterized matching for the forward string $T$
as in Algorithm~\ref{alg:pdawgmatching}.
Further, we obtain the following corollary that allows for
bidirectional parameterized pattern searches:

\begin{corollary}
  Using a pair of $\PDAWG(T)$ and $\PSTree(\rev{T})$ with
  pv-encodings $\lrangle{T}$ and $\lrangle{\rev{T}}$,
  one can perform forward and backward search to find all
  substrings of $T$ that p-match a given pattern $P$
  in $O(m \log(|\Pi|+|\Sigma|) + \occ)$ time, 
  where $m$ is the length of pattern $P$ and $\occ$ is the number of occurrences to report. 
\end{corollary}

\begin{proof}
For a query pattern $P$ that grows in both directions,
one can easily maintain $\langle P \rangle$
in $O(\log |\Pi|)$ time per added character using $O(|\Pi|)$ working space.
    
It is known that $\PSTree(\rev{T})$
allows for the navigation of a (reversed) pattern $\rev{P}$ of length $m$
in $O(m \log(|\Pi|+|\Sigma|))$ time~\cite{Baker93}.
This can be translated to an amortized $O(\log (|\Pi|+|\Sigma|))$-time
navigation per input character that is added to the \emph{left end}
of $P$ (which is the right-end of $\rev{P}$.)
A symmetric argument holds for our $\PDAWG(T)$,
which leads to an amortized $O(\log(|\Pi|+|\Sigma|))$-time
navigation per input character that is added to the \emph{right end} of $P$.

After locating the locus for the whole pattern $P$,
we can report all the $\occ$ occurrences in $O(\occ)$ time
using Theorem~\ref{thm:matching}.
\end{proof}

\subsection{Offline construction of PDAWGs via p-suffix trees}\label{sec:pdawg_pst}

In this section, we present a fast offline construction algorithm for $\PDAWG(T)$,
provided that $\PSTree(\rev{T})$ (without suffix links)
has already been built.

By the definition of our Weiner links on parameterized suffix trees, 
the following monotonicity holds.
\begin{lemma} \label{lem:monotonicity_Wlink}
	Suppose that a node $v$ in $\PSTree(\rev{T})$ has an (implicit or explicit)
	Weiner link $(v, k, u)$ with label $k \in \Sigma \cup \calN$.
	Then, any ancestor $v'$ of $v$ has an (implicit or explicit) Weiner link
        $(v', \Zinf{k}{|v'|}, u')$
        where $u'$ is the shallowest ancestor of $u$ with $|u'| \geq |v'|+1$.
\end{lemma}

It follows from Theorem~\ref{lem:duality} and Lemma~\ref{lem:monotonicity_Wlink}
that there is a simple \emph{offline} algorithm that builds the PDAWG
by computing Weiner links in a bottom-up manner over the PST
for the reversed text string.

\begin{theorem} \label{theo:offline_const}
        Let $T$ be a p-string of length $n$.
	Given $\PSTree(\rev{T})$ (without suffix links),
	we can build $\PDAWG(T)$ in $O(n)$ time and space.
\end{theorem}

\begin{figure}[t]
  \centering
  \includegraphics[scale=0.5]{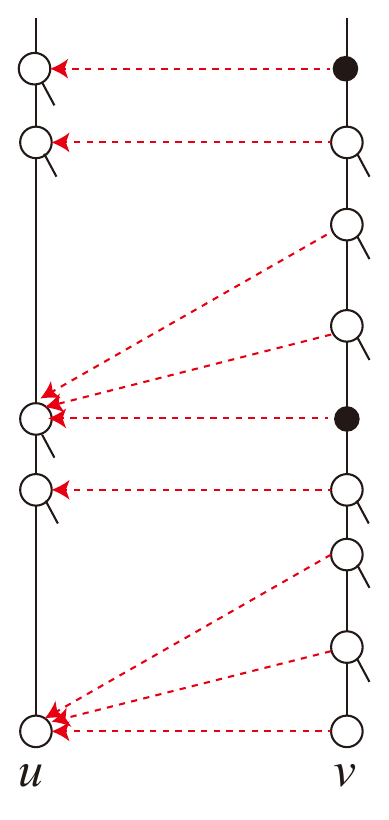}
  \caption{Illustration for our algorithm of Theorem~\ref{theo:offline_const} that propagates the Weiner links between the ancestors of $v$ and $u$, in a bottom up manner. The white circles represent nodes of $\PSTree(\rev{T})$ and the black circles represent imaginary nodes. The red arcs represent Weiner links.}
  \label{fig:Weiner_link_offline}
\end{figure}

\begin{figure}
  \centerline{
    \includegraphics[scale=0.25]{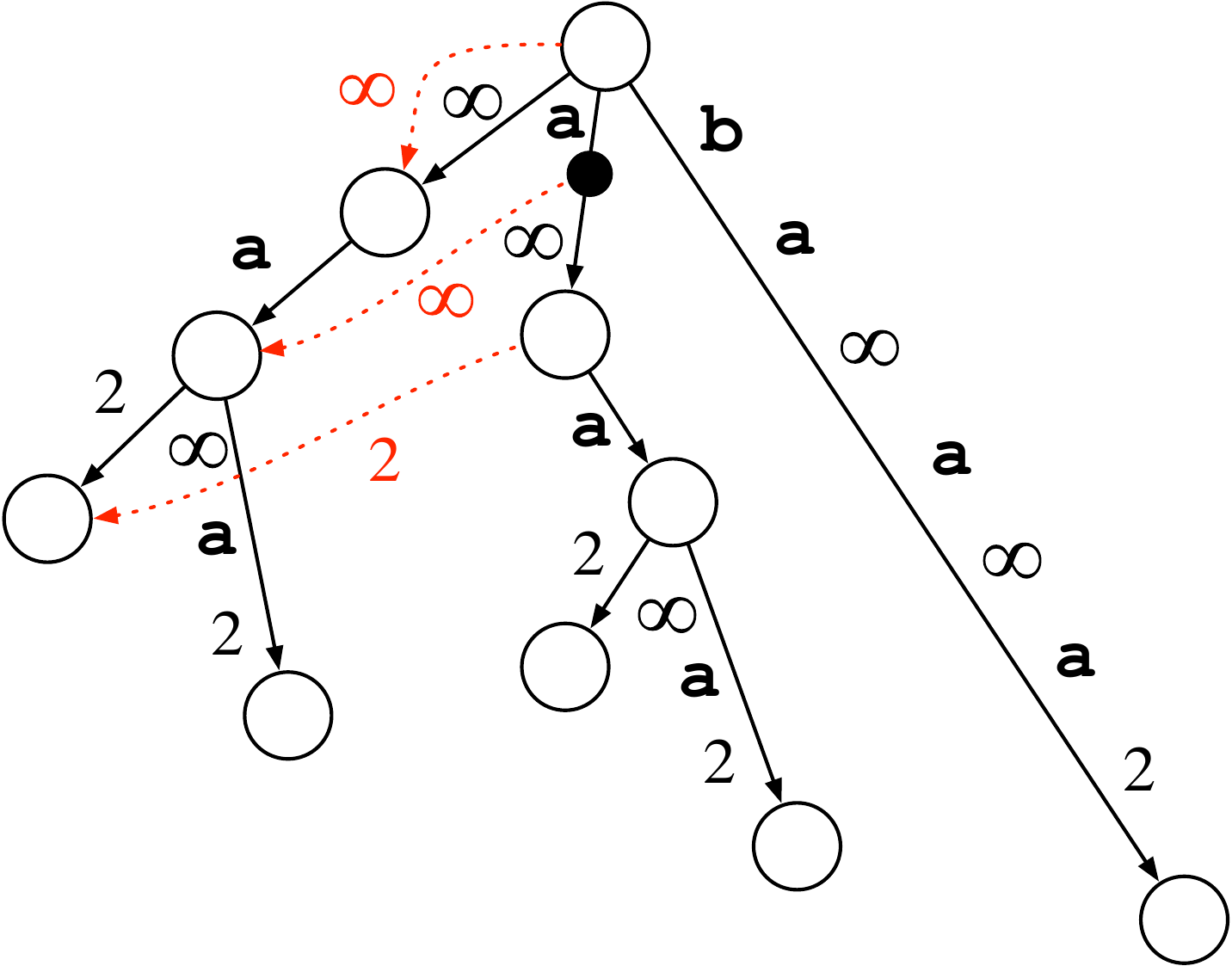}
  }
  \caption{An example for how the Weiner links are propagated in a bottom-up manner, with the same string as in Figure~\ref{fig:pSTree_pDAWG}. We pick the Weiner link $(\mathtt{a\inft}, {2}, \mathtt{\inft a}2)$ between two nodes $\mathtt{a\inft} = \lrangle{\mathtt{ay}}$ and $\mathtt{\inft a}2 = \lrangle{\mathtt{yay}}$. An imaginary node (black circle) is created at string depth $|\mathtt{\inft a}|-1 = 1$.
    By removing the imaginary nodes and the Weiner links from them, we obtain $\PDAWG(T)$ (see Figure~\ref{fig:pSTree_pDAWG}).
  }
  \label{fig:Weiner_link_offline_example}
\end{figure}

\begin{proof}
  We first compute the (reversed) suffix links of the nodes of $\PSTree(\rev{T})$ that corresponds to $\lrangle{S}$ for all suffixes of $T$, together with their labels which are the characters of $\lrangle{T}$. 

  For each Weiner link $L = (v, k, u)$ between two nodes
  corresponding to two consecutive suffixes of $T$, perform the following:
\begin{enumerate}
\item[(1)] Perform the following $\Update(v, u)$ function:
  \begin{enumerate}
  \item[(a)] If $|\parent(u)| = \parent(v)+1$, set $v \leftarrow \parent(v)$ and $u \leftarrow \parent(u)$.
  \item[(b)] If $|\parent(u)| < \parent(v)+1$, set $v \leftarrow \parent(v)$.
  \item[(c)] If $|\parent(u)| > \parent(v)+1$, then create an imaginary node $w$ at string depth $|\parent(u)|-1$ between $\parent(v)$ and $v$. Set $v \leftarrow w$ and $u \leftarrow \parent(u)$.
  \end{enumerate}
\item[(2)]
  \begin{enumerate}
    \item[(a)] If there is no Weiner link between $v$ and $u$, create a new Weiner link $(v, \Zinf{k}{|v|}, u)$. Go to (1).
    \item[(b)] Otherwise, stop the propagation for $L$.
  \end{enumerate}
\end{enumerate}
See also Figure~\ref{fig:Weiner_link_offline} that illustrates our algorithm.
The correctness of this algorithm is immediate from Lemma~\ref{lem:monotonicity_Wlink}.

It is clear from Lemma~\ref{lem:monotonicity_Wlink}
that the complexity of this algorithm is linear in the number
of Weiner links created.
It follows from Theorem~\ref{thm:pdawg_size} and Theorem~\ref{lem:duality}
that the number of Weiner links $(v, k, u)$
such that $v$ is \emph{not} an imaginary node is $O(n)$.
It also follows from our duality discussion in Section~\ref{sec:pdawg_pst}
and the definition of $\PPDAWG(T)$,
that each Weiner link $(w, k, u)$ such that $w$ is an imaginary node
corresponds to an edge in $E' \setminus E$,
where $E'$ is the set of edges of $\PPDAWG(T)$
and $E$ is the set of edges of $\PDAWG(T)$.
Since $|E'| = O(n)$ by Lemma~\ref{cor:ppdawg_edge_size},
the total time complexity of this algorithm is $O(n)$.
\end{proof}
See also Figure~\ref{fig:Weiner_link_offline_example}
for a concrete example of our offline algorithm
that computes $\PDAWG(T)$ and $\PPDAWG(T)$ from $\PSTree(\rev{T})$.

\section{Online algorithm for constructing PDAWGs}\label{sec:pdawg_construction}
This section proposes an algorithm constructing PDAWGs online.
Our algorithm is based on the one by Blumer et al.~\cite{DAWG} for constructing DAWGs of static strings.
In fact, if the input string is static, the behavior of our algorithm coincides with theirs.
In this section, we refer to $\PDAWG(T)$ as $\PDAWG(\lrangle{T})$ for $T \in (\Sigma \cup \Pi)^*$
and consider updating $\msf{PDAWG}(w)$ to $\msf{PDAWG}(wa)$ for a pv-string $wa$ where $a \in \Sigma \cup \calN$.
To distinguish sets of nodes, edges, and suffix links of $\msf{PDAWG}(w)$ and $\msf{PDAWG}(wa)$, we add subscripts $w$ and $wa$ to respective sets.

We first consider the difference of the node sets $V_w$ of $\PDAWG(w)$ and $V_{wa}$ of $\PDAWG(wa)$.
Recall that $V_w$ is a partition of $\PFactor(w)$.
Concerning strings in $\PFactor(wa)\setminus\PFactor(w) \subseteq \PSuffix(wa)$, we create a new sink node $\rec{wa}{wa} = \PFactor(wa)\setminus\PFactor(w)$ in $\PDAWG(wa)$ for which $\msf{RPos}_{wa}(\rec{wa}{wa}) = \{|wa|\}$.
Other strings $z$, which are already in $\PFactor(w)$, may or may not get a new end position $|wa|$, depending on whether they are in $\PSuffix(wa)$.
If all or no members of $\rec{z}{w}$ get the new position, we have $\rec{z}{w}=\rec{z}{wa}$. We keep those nodes.
However, it is possible that some but not all members of $\rec{z}{w}$ get the new end position.
In this case, the node $\rec{z}{w}$ will be split into two.
Let us define the \emph{longest repeated suffix (LRS)} of $wa \in (\Sigma \cup \calN)^+$ to be $\msf{LRS}(wa) = \longest{\PSuffix(wa) \cap \PFactor(w)}$.
Then, a pv-string $z \in \PFactor(w)$ will have $|wa| \in \msf{RPos}_{wa}(z)$ if and only if $z$ is a pv-suffix of $\msf{LRS}(wa)$.
That is, a node $\rec{z}{w}$ will be split if and only if some of $\rec{z}{w}$ are pv-suffix of $\msf{LRS}(wa)$ and some are not.
In other words, the split node in $V_w$ includes $\msf{LRS}(wa)$ and longer pv-strings.
On the other hand, if $\longest{\rec{\msf{LRS}(wa)}{w}} = \msf{LRS}(wa)$, then no node will be split in the update.
We call $\rec{\msf{LRS}(wa)}{w} \in V_w$ \emph{the LRS node} (w.r.t.\ $wa$).
The following lemma for node splits on PDAWGs is an analog to that for DAWGs.
\begin{lemma}[Node update]\label{lem:construction_nodes}
For $x = \msf{LRS}(wa)$ and $y = \longest{[x]^\mrm{R}_w}$,
\[
	V_{wa} = V_w \setminus \{[x]^\mrm{R}_{w}\} \cup \{[x]^\mrm{R}_{wa}, [y]^\mrm{R}_{wa}, [wa]^\mrm{R}_{wa}\}
\,.\]
If $x = y$, then $[x]^\mrm{R}_{w} = [x]^\mrm{R}_{wa} = [y]^\mrm{R}_{wa}$, i.e., $V_{wa} = V_w \cup \{[wa]^\mrm{R}_{wa}\}$.
Otherwise, $[x]^\mrm{R}_{w} = [x]^\mrm{R}_{wa} \cup [y]^\mrm{R}_{wa}$ and $[x]^\mrm{R}_{wa} \neq [y]^\mrm{R}_{wa}$.
\end{lemma}
\begin{proof}
	First remark that $\msf{RPos}_{wa}(z) = \msf{RPos}_{w}(z) \cup \{|wa|\}$ for all $z \in \PSuffix(wa)$ and $\msf{RPos}_{wa}(z) = \msf{RPos}_{w}(z) $ for all $z \notin \PSuffix(wa)$.
	For those $z \in \PSuffix(wa) \setminus \PFactor(w)$, we have $\msf{RPos}_{wa}(z) = \{|wa|\}$ and $[wa]^\mrm{R}_{wa} = \PSuffix(wa) \setminus \PFactor(w) \in V_{wa} \setminus V_w$.
	For $z \in \PFactor(w)$, if $[z]^\mrm{R}_w \neq [z]^\mrm{R}_{wa}$, some elements of $[z]^\mrm{R}_w$ are in $\PSuffix(wa)$ and some are not.
	That is, $[z]^\mrm{R}_w$ is partitioned into two non-empty equivalence classes 
	$\{\, z' \in [z]^\mrm{R}_w \mid z' \in \PSuffix(wa)\,\}$ and $\{\, z' \in [z]^\mrm{R}_w \mid z' \notin \PSuffix(wa)\,\}$.
	By definition, the longest of the former is $x=\msf{LRS}(wa)$ and the longest of the latter is $y=\longest{[x]^\mrm{R}_w}$.
	Otherwise, $[z]^\mrm{R}_w = [z]^\mrm{R}_{wa} \in V_w \cap V_{wa}$.
\end{proof}
\begin{example}[Figure~\ref{fig:incoming_edge}]
Let $w=\inft\mtt{a}2\mtt{a}$ and $a=\inft$.
Then, $\msf{LRS}(wa) = \lrangle{w[2:3]} = \lrangle{wa[4:5]} = \mtt{a}\inft$.
We have $\msf{LRS}(wa)\neq \longest{[\msf{LRS}(wa)]^\mrm{R}_w} = \inft\mtt{a}2$, where $\msf{RPos}_w(\mtt{a}\inft)=\msf{RPos}_w(\inft\mtt{a}2)=\{3\}$.
On the other hand, $\msf{RPos}_{wa}(\mtt{a}\inft)=\{3,5\} \neq \msf{RPos}_{wa}(\inft\mtt{a}2)=\{3\}$.
Therefore, $\PDAWG(wa)$ has two more nodes than $\PDAWG(w)$.
\end{example}
\begin{figure}[t]
	\definecolor{dgreen}{rgb}{0,0.5,0}
	\newcommand{\tcrd}[1]{\textcolor[rgb]{1.0,.0,.0}{#1}}
	\newcommand{\tcdg}[1]{\textcolor[rgb]{.0,.5,.0}{#1}}
	\newcommand{\LRSNode}{\LRS\LRSLong}
	\newcommand{\LRS}{$x$}
	\newcommand{\LRSLong}{$y$}
	\newcommand{\preLRS}{$x'$}
	\centering
	\begin{tikzpicture}[xscale=1,thick,shape=circle,inner sep=0pt,minimum size=4mm]
	\node[] (dummy) at (0,1) {};
	\node[draw] (e) at (0,0) {};
	\node[draw] (0) at (0,-1) {};
	\node[draw] (0a) at (0,-2) {};
	\node[draw] (0a2) at (0,-3) {};
	\node[draw] (0a2a) at (0,-4) {};
	\draw[edg] (dummy) -- (e);
	\draw[edg,out=190,in=170] (e) to node[left] {$\mtt{a}$} (0a);
	\draw[edg] (e) to node[right=-1] {$\inft$} (0);
	\draw[edg] (0) to node[right=-1] {$\mtt{a}$} (0a);
	\draw[edg] (0a) to node[right=-1] {$2$} (0a2);
	\draw[edg] (0a2) to node[right=-1] {$\mtt{a}$} (0a2a);
	\draw[sfl,out=150,in=215] (0) to (e);
	\draw[sfl,out=40,in=320] (0a2a) to (0a);
	\draw[sfl,out=0,in=0] (0a) to (e);
	\draw (0,-2) node {\small\preLRS};
	\draw (0,-3) node {\small\LRSNode};
	\draw[-latex,line width=3pt] (1.5,-2.25) -- (2.5,-2.25);
\begin{scope}[shift={(4.25,0)}]
	\node[] (dummy) at (0,1) {};
	\node[draw] (e) at (0,0) {};
	\node[draw] (a0) at (-1,-3) {};
	\node[draw] (0) at (0,-1) {};
	\node[draw] (0a) at (0,-2) {};
	\node[draw] (0a2) at (0,-3) {};
	\node[draw] (0a2a) at (0,-4) {};
	\node[draw] (0a2a0) at (0,-5) {};
	\draw[edg] (dummy) -- (e);
	\draw[edg,out=190,in=170] (e) to node[left] {$\mtt{a}$} (0a);
	\draw[edg,out=280,in=160] (a0) to node[below] {$\mtt{a}$} (0a2a);
	\draw[edg] (e) to node[right] {$\inft$} (0);
	\draw[edg] (0) to node[right] {$\mtt{a}$} (0a);
	\draw[edg,out=340,in=20] (0a) to node[right] {$\inft$} (0a2a0);
	\draw[edg] (0a) to node[right] {$2$} (0a2);
	\draw[edg] (0a2) to node[right] {$\mtt{a}$} (0a2a);
	\draw[edg] (0a2a) to node[right] {$\inft$} (0a2a0);
	\draw[sfl,out=160,in=270] (0a2a0) to (a0);
	\draw[sfl,out=90,in=215] (a0) to (0);
	\draw[sfl,out=150,in=215] (0) to (e);
	\draw[sfl,out=40,in=320] (0a2a) to (0a);
	\draw[sfl,out=0,in=0] (0a) to (e);
	\draw[sfl,out=180,in=0] (0a2) to (a0);
	\draw (0,-2) node {\small\preLRS};
	\draw (-1,-3) node {\small\LRS};
	\draw (0,-3) node {\small\LRSLong};
\end{scope}
	\end{tikzpicture}
	\caption{%
		$\PDAWG(w)$ and $\PDAWG(wa)$ for $w = \inft \mtt{a} 2 \mtt{a}$ and $a= \inft$.
		The strings
		$\msf{LRS}(wa)=x=\mtt{a}\inft$, $\msf{preLRS}(wa)= x' = \mtt{a}$, and $y = \longest{[x]^\mrm{R}_{w}}=\inft\mtt{a}2$
		are shown on the respective nodes where they belong.
	}
	\label{fig:incoming_edge}
\end{figure}

When updating $\msf{PDAWG}(w)$ to $\msf{PDAWG}(wa)$, all edges that do not touch the LRS node $[\msf{LRS}(wa)]^\mrm{R}_w$ are kept by definition.
What we have to do is to manipulate incoming edges for the new sink node $[wa]^\mrm{R}_{wa}$, and,
if necessary, to split the LRS node into two and to manipulate incoming and outgoing edges of them.
Therefore, it is very important to identify the LRS node $[\msf{LRS}(wa)]^\mrm{R}_w$ and to decide whether $\msf{LRS}(wa) = \longest{[\msf{LRS}(wa)]^\mrm{R}_w}$.
We remark that $\msf{LRS}(wa) = \varepsilon$ if and only if $wa \in \Sigma^*\{\inft\} \cup (\calN \cup \Sigma \setminus\{a\})^*\Sigma$.
This special case where $\msf{LRS}(wa) = \varepsilon$ is easy to handle, since the LRS node will never be split by $[\varepsilon]^\mrm{R}_w = \{\varepsilon\}$.
Hereafter, we assume that $\msf{LRS}(wa) \neq \varepsilon$ and define $\msf{preLRS}(wa)$ to be the prefix of $\msf{LRS}(wa)$ obtained by removing the last character: $\msf{preLRS}(wa) = \msf{LRS}(wa)[{}:|\msf{LRS}(wa)-1|]$.
Here, $\msf{LRS}(wa)$ and $\msf{preLRS}(wa)$ are pv-suffixes of $wa$ and $w$, respectively.
The LRS node can be reached from the pre-LRS node $[\msf{preLRS}(wa)]^\mrm{R}_w$ by reading one more character, and the pre-LRS node can be found by following suffix links from the sink node $[w]^\mrm{R}_w$ of $\msf{PDAWG}(w)$.
This appears quite similar to online construction of DAWGs for static strings, but there are nontrivial differences.
Main differences from the DAWG construction are in the following points:
\begin{itemize}
	\item Our PDAWG construction uses $\msf{trans}_w(u,i,\Zinf{a}{i})$ with an appropriate $i$, when the original DAWG construction refers to $\msf{child}_w(u,a)$,
	\item While $\msf{preLRS}(wa)$ is the longest of its equivalence class for static strings in $\msf{DAWG}(w)$, it is not necessarily the case for p-strings (like the one in Figure~\ref{fig:incoming_edge}).
	This affects the procedure to find the node of $\msf{LRS}(wa)$.
	As a consequence, if $\msf{preLRS}(wa) \neq \longest{\rec{\msf{preLRS}(wa)}{w}}$ and the LRS node is split, $\PDAWG(wa)$ has no edge from $\rec{\msf{preLRS}(wa)}{wa}$ to $\rec{\msf{LRS}(wa)}{wa}$.
	Even it is possible that $\rec{\msf{LRS}(wa)}{wa}$ has no incoming edges in $\PDAWG(wa)$.
	\item When the LRS node of $\msf{PDAWG}(w)$ is split into two in $\msf{PDAWG}(wa)$, 
	the outgoing edges of the two obtained nodes are identical in the DAWG construction, while it is not necessarily the case anymore in the PDAWG construction.
\end{itemize}

Let us call a sequence $\lrangle{v_0,v_1,\dots,v_k}$ of nodes of $V_w$ \emph{the suffix link chain of $v_0$} if $v_{i+1} = F_w(v_i)$ for $0 \le i < k$ and $v_k = \{\varepsilon\}$.
In DAWGs, the pre-LRS node is the first node with an $a$-edge on the suffix link chain of the old sink $[w]^\mrm{R}_w$.
However, it is not necessarily the case for PDAWGs.
Lemma~\ref{lem:prelrs} below suggests how to find the pre-LRS and LRS nodes, how to obtain the length of the LRS, and how to decide whether the LRS node should be split. 
The first item of the lemma describes how to find the pre-LRS node.
The pre-LRS node is on the suffix link chain $\lrangle{\rec{w}{w}=u_0,u_1,u_2,\dots,\rec{\varepsilon}{w}}$ of the old sink node $\rec{w}{w}$.
It is the firstly found node $u_j$ from which one can read $\Zinf{a}{ |\shortest{u_j}|}$.
The second item will be used to identify the length of the pre-LRS $x'$ and the LRS $x=x'a'$, where $a'=\Zinf{a}{|x'|}$.
If the pre-LRS node $u_j$ has an edge labeled with $\Zinf{a}{|\longest{u_j}|}$, then the longest element $\longest{u_j}$ is the pre-LRS $x'$ and thus $\lrangle{\longest{u_j}\cdot a} = \longest{u_j}\cdot \Zinf{a}{|\longest{u_j}|}$ is the LRS.
Otherwise, the lengths of the pre-LRS and the LRS can be determined by $a$ and the largest number (or $\inft$) labeling an outgoing edge of $u_j$. 
The third and fourth items are immediate consequences of Lemmas~\ref{lem:pdawgmatching} and~\ref{lem:construction_nodes}, respectively.
We can reach the LRS node $\rec{x'a'}{w}$ from the pre-LRS node by the transition function $\msf{trans}_w(u_j,|x'|,a')$.
The LRS node shall be split if and only if the LRS is not the longest of the node.
Hereafter, throughout this section, we fix the following variables:
$x = \msf{LRS}(wa)$,
$x' = \msf{preLRS}(wa)$, $a' = \Zinf{a}{|x'|}$, (i.e., $x=x'a'$,) $y=\longest{\rec{x}{w}}$, $u_0=\rec{w}{w}$, $u_i=F_w(u_{i-1})$ for $i \ge 1$  as long as $F_w(u_{i-1})$ is defined, and $a_i=\Zinf{a}{|\longest{u_i}|}$.

\begin{lemma}\label{lem:prelrs}
We have 
\begin{enumerate}
	\item $x' \in u_j$ for the least $j$ such that $\msf{trans}_w(u_j, |\shortest{u_j}|, \Zinf{a}{ |\shortest{u_j}|}) \neq \msf{Null}$,
	\item $|x'a'|= \begin{cases}
		|\longest{u_j}| + 1 \hfill \text{if\/ $\msf{child}_w(u_j, a_j) \neq \msf{Null}$,}
	\\	\min \{a,\,\max(\msf{Children}_w({u_j}) \cap \calN) \} \quad\quad\quad\quad\quad\quad \text{otherwise,}
	\end{cases}$\footnote{This corrects an error in the conference version~\cite{NakashimaFHNYIB20}.}
\\	for $j$ such that $x' \in u_j$,
	\item $[x'a']^\mrm{R}_w = \msf{trans}_w(u_j,|x'|,a')$ for $j$ such that $x' \in u_j$,
	\item $[x'a']^\mrm{R}_w \neq [x'a']^\mrm{R}_{wa}$ if and only if\/ $|x'a'| \neq |\longest{[x'a']^\mrm{R}_w}|$.
	\end{enumerate}
\end{lemma}
\begin{proof}
	Suppose $x' \in u_j$. Note that $\PSuffix(w) = \bigcup_{i \ge 0} u_i$.
	
	(1)
	 Since every string $z \in u_i \subseteq \PSuffix(w)$ with $i<j$ is properly longer than $x'$, particularly for $z=\shortest{u_i}$, $\lrangle{za} \in \PSuffix(wa)$ is properly longer than the LRS $x=x'a'$.
	Thus,  $\lrangle{za}  = z \Zinf{a}{|z|} \notin \PFactor(w)$.
	Therefore, $\msf{trans}_w(u_i, |\shortest{u_i}|, \Zinf{a}{|\shortest{u_i}|}) = \msf{Null}$.
	On the other hand, the fact $\shortest{u_j} \in \PSuffix(x')$ implies $\lrangle{\shortest{u_j} a} = \shortest{u_j} \Zinf{a}{|\shortest{u_j}|} \in \PSuffix(x'a') \subseteq \PFactor(w)$.
	Hence, $\msf{trans}_w(u_j, |\shortest{u_j}|, \Zinf{a}{|\shortest{u_j}|}) \neq \msf{Null}$.
	
	(2) 
	Recall that for any element $z$ of the pre-LRS node $u_j$, $\lrangle{za} \in \PFactor(w)$ if and only if $|\lrangle{za}| \le |x'a'|$. 
	Hence, $\msf{child}_w(u_j, a_j) \neq \msf{Null}$ if and only if  the longest element $\longest{u_j}$ of the pre-LRS node $u_j$ is the pre-LRS, i.e., $|x'a'|=|\longest{u_j}|+1$.

	Now suppose $\msf{child}_w(u_j, a_j) = \msf{Null}$, but $\msf{trans}_w(u_j, |\shortest{u_j}|, \Zinf{a}{ |\shortest{u_j}|}) \neq \msf{Null}$.
	Then, $\longest{u_j}$ is properly longer than $x'$.
	Let $x'' \in u_j$ be the element of $u_j$ of length $|x'|+1$ and $a'' = \Zinf{a}{|x''|}$.
	Since $x'a'$ is the LRS, the longer p-suffix $x''a'' \in \PSuffix(wa)$ does not occur in $w$, so $x'a' \not\equiv_w^\mrm{R} x''a''$, whereas $x' \equiv_w^\mrm{R} x''$.
	Here, one can see that $a' = \inft$ as follows.
	If $a' = a''$, then $a' = \inft$ by Lemma~\ref{lem:ppdawg1}.
	If $a' \neq a''$, then it can happen only when $|x'| < a'' = a < \inft$ and $a' = \inft$.
	Let $Z = \msf{Children}_w({u_j})\cap \calN$ and $Z_i = \{\, k \in Z \mid k > i \,\}$.
	Lemma~\ref{lem:pdawgmatching} and the fact $x'a' \in \PFactor(w)$ imply $Z_{|x'|} \neq \emptyset$.
	On the other hand, $x''a'' \notin \PFactor(w)$ implies
	 either $a''=\inft$ and $Z_{|x''|} = \emptyset$ or $0 < a''=a < \inft$.
	In the former case, $Z_{|x'|} \neq Z_{|x''|} = \emptyset$ implies that $Z_{|x'|}$ is the singleton set with the element $\max Z = |x''|$. 
	By $a'=\Zinf{a}{|x'|}=\inft$, we have $|x'| < a$.
	That is, $\max Z = |x''| = |x'a'| \le a$.
	In the latter case, $a' = \Zinf{a}{|x'|} = \inft \neq a'' = \Zinf{a}{|x''|} \in \mbb{N}$ implies $a = |x''|$.
	By $Z_{|x'|} \neq \emptyset$, we have $|x'| < \max Z$.
	That is,  $a = |x''| = |x'a'| \le \max Z$.
	Summarizing these cases, $|x'a'| = \min \{a,\,\max Z \}$.
	
	(3) By Lemma~\ref{lem:pdawgmatching}.
	(4) By Lemma~\ref{lem:construction_nodes}.
\end{proof}
To update the PDAWG based on Lemma~\ref{lem:prelrs}, we need to know the lengths of the longest and shortest elements of each node.
Accordingly, our algorithm maintains the value $\msf{len}(u)=|\longest{u}|$ for each node $u$.
The length of the shortest element can be calculated by $|\shortest{u}|=|\longest{F(u)}|+1= \msf{len}(F(u))+1$.

\begin{example}
See Figure~\ref{fig:incoming_edge}, where $wa = \inft \mtt{a} 2 \mtt{a} \inft$, and the pre-LRS and LRS are $x'=\mtt{a}$ and $x=\mtt{a}\inft$, respectively.
The pre-LRS node $\rec{x'}{w}$ is the first node $u$ on the suffix link chain of the old sink $\rec{w}{w}$ such that $\msf{trans}_w(u,\shortest{u},\inft) \neq \msf{Null}$.
This is how we find the pre-LRS node by Lemma~\ref{lem:prelrs}.1.
The lengths of the pre-LRS and LRS can be known by Lemma~\ref{lem:prelrs}.2.
In this case, $\msf{child}_w(\rec{x'}{w},\inft) = \msf{Null}$.
Hence, $|x| = \min\{\inft,\max \msf{Children}_w(\rec{x'}{w})\cap\calN\} = 2$ and thus $|x'|=1$.
From the identified pre-LRS node, one can reach the LRS node by $\msf{trans}_w(\rec{x'}{w},1,\inft) = \rec{x}{w}$.
Here, $|\longest{ \rec{x}{w}}|=3 \neq |x| = 2$.
So, the LRS node should be split.

Figure~\ref{fig:ltor_example} shows another example, with a step-by-step illustration of our algorithm, which will be explained in more detail later.
Compare the initial PDAWG (a) and the goal PDAWG (d).
For $wa = \inft \mtt{a} 2 \mtt{a} \inft \mtt{aa}$, the pre-LRS and LRS are $x'=\varepsilon$ and $x=\mtt{a}$, respectively.
Indeed, the source node $\rec{\varepsilon}{w}$ is the only node on the suffix link chain of the old sink $\rec{w}{w}$ that has an $\mtt{a}$-edge.
Since $\msf{child}_w(\rec{\varepsilon}{w},\mtt{a}) \neq \msf{Null}$, we know $|x|=|\longest{\rec{\varepsilon}{w}}|+1=1$ and $|x'|=0$.
The LRS node can be found by $\msf{trans}_w(\rec{x'}{w},0,\mtt{a}) = \rec{x}{w}$.
Since $|\longest{\rec{x}{w}}| = 2 > |x|=1$, the LRS node should be split.
\end{example}

\begin{figure}[!t]
	\definecolor{dgreen}{rgb}{0,0.5,0}
	\newcommand{\tcrd}[1]{\textcolor[rgb]{1.0,.0,.0}{#1}}
	\newcommand{\tcdg}[1]{\textcolor[rgb]{.0,.5,.0}{#1}}
	\newcommand{\LRSNode}{\LRS\LRSLong}
	\newcommand{\LRS}{$x$}
	\newcommand{\LRSLong}{$y$}
	\newcommand{\preLRS}{$x'$}
	\centering
	\begin{tikzpicture}[xscale=1,yscale=.95,thick,shape=circle,inner sep=0pt,minimum size=4mm]\small
\begin{scope}[shift={(0,0)}]
	\draw (0,2) node{$\PDAWG(\inft\mtt{a}2\mtt{a}\inft\mtt{a})$};	
	\node[dummy] (dummy) at (0,1) {};
	\node[draw] (e) at (0,0) {};
	\node[draw] (a0) at (-1,-3) {};
	\node[draw] (a0a) at (-1,-4) {};
	\node[draw] (0) at (0,-1) {};
	\node[draw] (0a) at (0,-2) {};
	\node[draw] (0a2) at (0,-3) {};
	\node[draw] (0a2a) at (0,-4) {};
	\node[draw] (0a2a0) at (0,-5) {};
	\node[draw] (0a2a0a) at (0,-6) {};
	\draw[edg] (dummy) to node[left] {\small$\Sigma\!\cup\!\{\inft\}$} (e);
	\draw[edg] (e) to [out=190,in=170] node[left=-1] {$\mtt{a}$} (0a);
	\draw[edg] (a0) to node[left=-1] {$\mtt{a}$} (a0a);
	\draw[edg] (a0a) to node[below] {$\inft$} (0a2a0);
	\draw[edg] (e) to node[right=-1] {$\inft$} (0);
	\draw[edg] (0) to node[right=-1] {$\mtt{a}$} (0a);
	\draw[edg] (0a) to [out=340,in=90] (0.7,-3.5) node[right=-2] {$\inft$} to [out=270,in=20] (0a2a0);
	\draw[edg] (0a) to node[right=-1] {$2$} (0a2);
	\draw[edg] (0a2) to node[right=-1] {$\mtt{a}$} (0a2a);
	\draw[edg] (0a2a) to node[right=-1] {$\inft$} (0a2a0);
	\draw[edg] (0a2a0) to node[right=-1] {$\mtt{a}$} (0a2a0a);
	\draw[sfl,out=135,in=290] (0a2a0a) to (a0a);
	\draw[sfl,out=120,in=315] (0a2a0) to (a0);
	\draw[sfl,out=70,in=225] (a0) to (0);
	\draw[sfl,out=150,in=215] (0) to (e);
	\draw[sfl] (0a2a) to (a0a);
	\draw[sfl,out=45,in=240] (a0a) to (0a);
	\draw[sfl,out=0,in=0] (0a) to (e);
	\draw[sfl,out=180,in=0] (0a2) to (a0);
	\draw[sfl,out=30,in=330] (e) to (dummy);
	\draw[sfl,out=30,in=90,loop] (dummy) to (dummy);
	\draw(0,0) node{{\small\preLRS}};
	\draw(0,-2) node{{\small\LRSNode}};
	\draw(0,-7.8) node{(a)};
\end{scope}
\begin{scope}[shift={(2.5,0)}]
	\node[dummy] (dummy) at (0,1) {};
	\node[draw] (e) at (0,0) {};
	\node[draw] (a0) at (-1,-3) {};
	\node[draw] (a0a) at (-1,-4) {};
	\node[draw] (0) at (0,-1) {};
	\node[draw] (0a) at (0,-2) {};
	\node[draw] (0a2) at (0,-3) {};
	\node[draw] (0a2a) at (0,-4) {};
	\node[draw] (0a2a0) at (0,-5) {};
	\node[draw] (0a2a0a) at (0,-6) {};
	\node[draw,red] (0a2a0aa) at (0,-7) {};
	\draw[edg] (dummy) to node[left=-1] {\small$\Sigma\!\cup\!\{\inft\}$} (e);
	\draw[edg] (e) to [out=190,in=170] node[left=-1] {$\mtt{a}$} (0a);
	\draw[edg] (a0) to node[left=-1] {$\mtt{a}$} (a0a);
	\draw[edg] (a0a) to node[below] {$\inft$} (0a2a0);
	\draw[edg] (e) to node[right=-1] {$\inft$} (0);
	\draw[edg] (0) to node[right=-1] {$\mtt{a}$} (0a);
	\draw[edg] (0a) to [out=315,in=90] (0.5,-3.5) node[right=-2] {$\inft$} to [out=270,in=45] (0a2a0);
	\draw[edg] (0a) to node[right=-1] {$2$} (0a2);
	\draw[edg] (0a2) to node[right=-1] {$\mtt{a}$} (0a2a);
	\draw[edg] (0a2a) to node[right=-1] {$\inft$} (0a2a0);
	\draw[edg] (0a2a0) to node[right=-1] {$\mtt{a}$} (0a2a0a);
	\draw[edg,red] (0a2a0a) to node[right=-1] {$\mtt{a}$} (0a2a0aa);
	\draw[edg,red] (a0a) to [out=270,in=135] node[left=-1] {$\mtt{a}$} (0a2a0aa);
	\draw[edg,red] (0a) to [out=340,in=90] (1.0,-4.5) node[left=-2] {$\mtt{a}$} to [out=270,in=20] (0a2a0aa);
	\draw[sfl,dgreen,out=135,in=290] (0a2a0a) to (a0a);
	\draw[sfl,out=120,in=315] (0a2a0) to (a0);
	\draw[sfl,out=70,in=225] (a0) to (0);
	\draw[sfl,out=150,in=215] (0) to (e);
	\draw[sfl] (0a2a) to (a0a);
	\draw[sfl,dgreen,out=45,in=240] (a0a) to (0a);
	\draw[sfl,dgreen,out=0,in=0] (0a) to (e);
	\draw[sfl,out=180,in=0] (0a2) to (a0);
	\draw[sfl,out=30,in=330] (e) to (dummy);
	\draw[sfl,out=30,in=90,loop] (dummy) to (dummy);
	\draw(0,0) node{{\small\preLRS}};
	\draw(0,-2) node{{\small\LRSNode}};
	\draw(0,-7.8) node{(b)};
\end{scope}
\begin{scope}[shift={(5.75,0)}]
	\node[dummy] (dummy) at (0,1) {};
	\node[draw] (e) at (0,0) {};
	\node[draw] (a0) at (-1,-3) {};
	\node[draw] (a0a) at (-1,-4) {};
	\node[draw] (0) at (0,-1) {};
	\node[draw] (0a) at (0,-2) {};
	\node[draw] (0a2) at (0,-3) {};
	\node[draw] (0a2a) at (0,-4) {};
	\node[draw] (0a2a0) at (0,-5) {};
	\node[draw] (0a2a0a) at (0,-6) {};
	\node[draw] (0a2a0aa) at (0,-7) {};
	\node[draw,red] (a) at (-1,-2) {};
	\draw[edg] (dummy) to node[left=-1] {\small$\Sigma\!\cup\!\{\inft\}$} (e);
	\draw[edg,red] (e) to [out=180,in=90] node[left] {$\mtt{a}$} (a);
	\draw[edg] (a0) to node[left=-1] {$\mtt{a}$} (a0a);
	\draw[edg] (a0a) to node[below] {$\inft$} (0a2a0);
	\draw[edg] (e) to node[right=-1] {$\inft$} (0);
	\draw[edg] (0) to node[right=-1] {$\mtt{a}$} (0a);
	\draw[edg] (0a) to [out=315,in=90] (0.5,-3.5) node[right=-2] {$\inft$} to [out=270,in=45] (0a2a0);
	\draw[edg,dgreen] (0a) to node[right=-1] {$2$} (0a2);
	\draw[edg] (0a2) to node[right=-1] {$\mtt{a}$} (0a2a);
	\draw[edg] (0a2a) to node[right=-1] {$\inft$} (0a2a0);
	\draw[edg] (0a2a0) to node[right=-1] {$\mtt{a}$} (0a2a0a);
	\draw[edg] (0a2a0a) to node[right=-1] {$\mtt{a}$} (0a2a0aa);
	\draw[edg] (a0a) to [out=270,in=135] node[left=-1] {$\mtt{a}$} (0a2a0aa);
	\draw[edg,dgreen] (0a) to [out=340,in=90] (1.0,-4.5) node[left=-2] {$\mtt{a}$} to [out=270,in=20] (0a2a0aa);
	\draw[edg,red] (a) to [out=225,in=90] (-1.6,-4.5) node[right=-2] {$\mtt{a}$} to [out=270,in=150] (0a2a0aa);
	\draw[edg,red] (a) to node[left=-1] {$\inft$} (a0);
	\draw[sfl,out=135,in=290] (0a2a0a) to (a0a);
	\draw[sfl,out=120,in=315] (0a2a0) to (a0);
	\draw[sfl,out=70,in=225] (a0) to (0);
	\draw[sfl,out=150,in=215] (0) to (e);
	\draw[sfl] (0a2a) to (a0a);
	\draw[sfl,out=45,in=240] (a0a) to (0a);
	\draw[sfl,out=0,in=0] (0a) to (e);
	\draw[sfl,dgreen,out=180,in=0] (0a2) to (a0);
	\draw[sfl,out=30,in=330] (e) to (dummy);
	\draw[sfl,out=30,in=90,loop] (dummy) to (dummy);
	\draw(0,0) node{{\small\preLRS}};
	\draw(0,-2) node{{\small\LRSLong}};
	\draw(-1,-2) node{{\small\LRS}};
	\draw(0,-7.8) node{(c)};
\end{scope}
\begin{scope}[shift={(9.25,0)}]
	\draw (-0.25,2) node{$\PDAWG(\inft\mtt{a}2\mtt{a}\inft\mtt{aa})$};	
	\node[dummy] (dummy) at (0,1) {};
	\node[draw] (e) at (0,0) {};
	\node[draw] (a0) at (-1,-3) {};
	\node[draw] (a0a) at (-1,-4) {};
	\node[draw] (0) at (0,-1) {};
	\node[draw] (0a) at (0,-2) {};
	\node[draw] (0a2) at (0,-3) {};
	\node[draw] (0a2a) at (0,-4) {};
	\node[draw] (0a2a0) at (0,-5) {};
	\node[draw] (0a2a0a) at (0,-6) {};
	\node[draw] (0a2a0aa) at (0,-7) {};
	\node[draw] (a) at (-1,-2) {};
	\draw[edg] (dummy) to node[left=-1] {\small$\Sigma\!\cup\!\{\inft\}$} (e);
	\draw[edg] (e) to [out=180,in=90] node[left] {$\mtt{a}$} (a);
	\draw[edg] (a0) to node[left=-1] {$\mtt{a}$} (a0a);
	\draw[edg] (a0a) to node[below] {$\inft$} (0a2a0);
	\draw[edg] (e) to node[right=-1] {$\inft$} (0);
	\draw[edg] (0) to node[right=-1] {$\mtt{a}$} (0a);
	\draw[edg] (0a) to [out=315,in=90] (0.5,-3.5) node[right=-2] {$\inft$} to [out=270,in=45] (0a2a0);
	\draw[edg] (0a) to node[right=-1] {$2$} (0a2);
	\draw[edg] (0a2) to node[right=-1] {$\mtt{a}$} (0a2a);
	\draw[edg] (0a2a) to node[right=-1] {$\inft$} (0a2a0);
	\draw[edg] (0a2a0) to node[right=-1] {$\mtt{a}$} (0a2a0a);
	\draw[edg] (0a2a0a) to node[right=-1] {$\mtt{a}$} (0a2a0aa);
	\draw[edg] (a0a) to [out=270,in=135] node[left=-1] {$\mtt{a}$} (0a2a0aa);
	\draw[edg] (0a) to [out=340,in=90] (1.0,-4.5) node[left=-2] {$\mtt{a}$} to [out=270,in=20] (0a2a0aa);
	\draw[edg] (a) to [out=225,in=90] (-1.6,-4.5) node[right=-2] {$\mtt{a}$} to [out=270,in=150] (0a2a0aa);
	\draw[edg] (a) to node[left=-1] {$\inft$} (a0);
	\draw[sfl,red] (0a2a0aa) to [out=180,in=270] (-2,-4.5) to [out=90,in=195] (a);
	\draw[sfl,out=135,in=290] (0a2a0a) to (a0a);
	\draw[sfl,out=120,in=315] (0a2a0) to (a0);
	\draw[sfl,out=60,in=240] (a0) to (0);
	\draw[sfl,out=30,in=330] (0) to (e);
	\draw[sfl] (0a2a) to (a0a);
	\draw[sfl,out=45,in=240] (a0a) to (0a);
	\draw[sfl,red] (0a) to (a);
	\draw[sfl,out=180,in=0] (0a2) to (a0);
	\draw[sfl,red] (a) to (e);
	\draw[sfl,out=30,in=330] (e) to (dummy);
	\draw[sfl,out=30,in=90,loop] (dummy) to (dummy);
	\draw(0,0) node{{\small\preLRS}};
	\draw(0,-2) node{{\small\LRSLong}};
	\draw(-1,-2) node{{\small\LRS}};
	\draw(0,-7.8) node{(d)};
\end{scope}
\end{tikzpicture}
	\caption{%
		Updating $\PDAWG(w)$ (a) to $\PDAWG(wa)$ (d) for $w = \lrangle{\mtt{xaxaya}}$ and $wa = \lrangle{\mtt{xaxayaa}}$.
		The subfigures (b) and (c) illustrate the structures right after Lines~\ref{algln:sinkedges} and~\ref{algln:incoming2} of Algorithm~\ref{alg:construction}, respectively.
		The circle with a broken line represents the dummy node $\top$.
		We present $x=\msf{LRS}(wa)=\mtt{a}$, $y=\longest{[\msf{LRS}(wa)]^\mrm{R}_{w}}=\inft\mtt{a}$, and $x'=\msf{preLRS}(wa)=\varepsilon$ in the respective nodes where they belong.
		Algorithm~\ref{alg:construction} modifies (a) into (b) by creating the new sink, adding incoming $\mtt{a}$-edges to the new sink from the nodes reachable by suffix links from the old sink, until arriving at a node that already has an $\mtt{a}$-edge.
		This node is the pre-LRS node $[x']_w^\mrm{R}$. 
		In (b), the newly added node and edges are colored red and the suffix links we followed are green.
		We now know the lengths of the pre-LRS and the LRS are $0$ and $1$, respectively.
		The LRS node $v$ is $\msf{trans}([x']_w^\mrm{R},0,\mtt{a})=[x]_w^\mrm{R}$.
		Since $\msf{len}(v) > 1$, the LRS node shall be split into $\rec{y}{wa}$ and $\rec{x}{wa}$, as shown in (c).
		Our algorithm recycles the LRS node $v$ for $\rec{y}{wa}$ and create $v'$ for $\rec{x}{wa}$.
		Then, the outgoing $\mtt{a}$-edge of the pre-LRS node is redirected from $v$ to $v'$.
		The $\mtt{a}$-edge of $v$ is copied for $v'$, which has just been added in the previous step in this particular example.
		In addition, $v'$ gets an $\inft$-edge toward $\msf{trans}_w(v,1,\inft)$.
		The new node $v'$ and its incoming and outgoing edges are colored red,
		and the edges and the suffix link referenced for making those outgoing edges are colored green in (c).    
		Finally, we obtain (d) by creating suffix links from the new sink to the new node, and from the new node $v'$ to the node that the old suffix link of $v$ points at, and by redirecting that of $v$ toward $v'$. Those suffix links are colored red.
	}
	\label{fig:ltor_example}
\end{figure}

Edges are created or replaced in accordance with the definition of a PDAWG.
The incoming edges for the new sink node $[wa]^\mrm{R}_{wa}$ of $\msf{PDAWG}(wa)$ are given as Lemma~\ref{lem:construction_sink_edges},
except for when we have an edge from $\rec{\msf{LRS}(wa)}{wa}$ to the new sink in the case of node split.
This case will be discussed later in Lemma~\ref{lem:construction_outgoing_edges}.
\begin{lemma}[Incoming edges of the new sink]\label{lem:construction_sink_edges}
	Let $u_i'= \rec{\longest{u_i}}{wa}$, which coincides with $u_i$ unless $u_i$ is the LRS node and split.
	The incoming edges for the new sink $[wa]^\mrm{R}_{wa}$ are exactly those $(u_i',a_i,[wa]^\mrm{R}_{wa})$ such that\/
	 $\msf{child}_w(u_i,a_i)=\msf{Null}$, except for an edge from $\rec{\msf{LRS}(wa)}{wa}$ in the case of node split.
\end{lemma}
\begin{proof}
	Recall that $\msf{RPos}_{wa}(wa)=\{|wa|\}$.
	For $[wa]^\mrm{R}_{wa}$ to have an incoming edge from a node $u \in V_{wa} \cap V_w$, $\longest{u}$ must occur as a pv-suffix of $w$.
	Then, the node $u$ must be in the suffix link chain of the old sink node, i.e., $u = u_i$ for some $i$.
	Here, $\longest{u_i}a_i \in \PSuffix(wa)$. 
	On the other hand, $\longest{u_i}a_i$ should not occur anywhere in $w$, since it occurs only at $|wa|$ in $wa$.
	That is, $u_i$ should not have an edge $a_i$ in $\PDAWG(w)$.
	Therefore, we must have $(u_i,a_i,\rec{wa}{wa}) \in E_{wa}$ if and only if $\msf{child}_w(u_i,a_i)=\msf{Null}$.
	
	It remains to consider whether we should have $(\rec{y}{wa},\Zinf{a}{|y|},\rec{wa}{wa}) \in E_{wa}$ in the case of node split.
	This can be discussed in essentially the same way as above.
	We have $(\rec{y}{wa},\Zinf{a}{|y|},\rec{wa}{wa}) \in E_{wa}$ if and only if $\rec{y}{w} = u_i$, i.e., $\rec{y}{wa}=u_i'$, for some $i$ and $\msf{child}_w(u_i,a_i)=\msf{Null}$.
\end{proof}
This is not much different from DAWG update, except that the pre-LRS node has an edge toward the new sink when the pre-LRS is not the longest in the pre-LRS node.
\begin{example}
In Figure~\ref{fig:incoming_edge}, the new sink gets an incoming edge from the pre-LRS node $[x']^\mrm{R}_{wa}$ in addition to the one from the old sink.
In Figure~\ref{fig:ltor_example}, the new sink gets incoming edges from the nodes on the suffix link chain of the old sink located \emph{before} the pre-LRS node, including the one from $\rec{y}{wa}$.
\end{example}

If the LRS node $[\msf{LRS}(wa)]^\mrm{R}_{wa}$ is not split, we have nothing more to do on edges.

Hereafter, we suppose that the LRS node shall be split.
That is, $x \neq y$ for $x = \msf{LRS}(wa)$ and $y =\longest{[\msf{LRS}(wa)]^\mrm{R}_w}$.
By definition, all edges of $\msf{PDAWG}(w)$ that do not involve the LRS node $[\msf{LRS}(wa)]^\mrm{R}_w$ will be inherited to $\msf{PDAWG}(wa)$.
First, we make it clear that we will have no edge between $\rec{x}{wa}$ and $\rec{y}{wa}$.
To have $(\rec{x}{wa},b,\rec{y}{wa}) \in E_{wa}$ for some $b \in \Sigma \cup \calN$, it requires $\msf{RPos}_{wa}(xb) = \msf{RPos}_{wa}(y)$.
Particularly for $i = \min\msf{RPos}_{wa}(xb)$, we have $i - 1 \in \msf{RPos}_{wa}(x)$, which contradicts $i = \min\msf{RPos}_{wa}(y)= \min\msf{RPos}_{w}(y)= \min\msf{RPos}_{w}(x)=\min\msf{RPos}_{wa}(x)$.
Similarly we conclude $(\rec{y}{wa},b,\rec{x}{wa}) \notin E_{wa}$.

Lemma~\ref{lem:construction_outgoing_edges} below is concerned with the outgoing edges of $\rec{y}{wa}$ and $\rec{x}{wa}$.
In the DAWG construction, those two nodes $[x]^\mrm{R}_{wa}$ and $[y]^\mrm{R}_{wa}$ simply inherit the outgoing edges of the LRS node $[x]^\mrm{R}_w =[y]^\mrm{R}_w$.
However, in the PDAWG construction, due to the prev-encoding rule on parameter characters, the node $[x]^\mrm{R}_{wa}$ will lose edges whose labels are integers greater than $|x|$, as demonstrated in Figure~\ref{fig:outgoing_edge}.
Those edges are ``bundled'' into a single $\inft$-edge which points at $\msf{trans}_w([y]^\mrm{R}_{w},|x|,\inft)$.
This is described as the second item of Lemma~\ref{lem:construction_outgoing_edges}.
\begin{figure}[t]
	\centering
	\begin{tikzpicture}[xscale=1.1,thick,shape=circle,inner sep=0pt,minimum size=6.5mm]\small
	\node[draw] (00a) at (3,0) {\scriptsize$\inft\inft\mtt{a}$};
	\node[draw] (00a0) at (4,-2) {\scriptsize$\inft\inft\mtt{a}\inft$};
	\node[draw] (00a3) at (3,-2) {\scriptsize$\inft\inft\mtt{a}3$};
	\node[draw] (00a2) at (2,-2) {\scriptsize$\inft\inft\mtt{a}2$};
	\node[draw] (0a0) at (1,-2) {\scriptsize$\inft\mtt{a}\inft$};
	\node[draw] (a0) at (0,-2) {\scriptsize$\mtt{a}\inft$};
	\draw[edg,out=315,in=100] (00a) to node[right=-5] {$\inft$} (00a0);
	\draw[edg] (00a) to node[right=-5] {$3$} (00a3);
	\draw[edg,out=225,in=80] (00a) to node[right=-5] {$2$} (00a2);
	\draw[sfl,out=240,in=300] (00a0) to (0a0);
	\draw[sfl,out=225,in=315] (00a3) to (0a0);
	\draw[sfl,out=225,in=315] (00a2) to (a0);
	\draw[sfl] (0a0) to (a0);
	\draw[-latex,line width=3pt] (4.6,-1) -- (5.6,-1);
\begin{scope}[shift={(6,0)}]
	\node[draw] (00a) at (3,0) {\scriptsize$\inft\inft\mtt{a}$};
	\node[draw] (00a0) at (4,-2) {\scriptsize$\inft\inft\mtt{a}\inft$};
	\node[draw] (00a3) at (3,-2) {\scriptsize$\inft\inft\mtt{a}3$};
	\node[draw] (00a2) at (2,-2) {\scriptsize$\inft\inft\mtt{a}2$};
	\node[draw] (0a0) at (1,-2) {\scriptsize$\inft\mtt{a}\inft$};
	\node[draw] (a0) at (0,-2) {\scriptsize$\mtt{a}\inft$};
	\node[draw] (0a) at (1,0) {\scriptsize$\inft\mtt{a}$};
	\draw[edg,out=315,in=100] (00a) to node[right=-5] {$\inft$} (00a0);
	\draw[edg] (00a) to node[right=-5] {$3$} (00a3);
	\draw[edg,out=225,in=80] (00a) to node[right=-5] {$2$} (00a2);
	\draw[edg,out=315,in=100] (0a) to node[left=-5] {$2$} (00a2);
	\draw[edg] (0a) to node[left=-5] {$\inft$} (0a0);
	\draw[sfl,out=240,in=300] (00a0) to (0a0);
	\draw[sfl,out=225,in=315] (00a3) to (0a0);
	\draw[sfl,out=225,in=315] (00a2) to (a0);
	\draw[sfl] (0a0) to (a0);
	\draw[sfl] (00a) to (0a);
\end{scope}
	\end{tikzpicture}
		\caption{\label{fig:outgoing_edge}
			Subgraphs of $\PDAWG(w)$ and $\PDAWG(wa)$ for $w= \inft\inft \mtt{a} 3\inft \mtt{a} 2\inft \mtt{a}\inft$ and $a= \mtt{a}$,
			where $\msf{LRS}(wa)=\inft\mtt{a}$ and $\longest{[\inft\mtt{a}]^\mrm{R}_w}=\inft\inft\mtt{a}$.
			For explanation, we show $\longest{u}$ in each node $u$.
			$[\inft \mtt{a}]_{wa}^\mrm{R}$ does not inherit the outgoing edges of $[\inft \mtt{a}]_{w}^\mrm{R}=[\inft\inft \mtt{a}]_{w}^\mrm{R}$ labeled with $3$ and $\inft$.
			Instead, the $3$-edge and $\inft$-edge are bundled into a single $\inft$-edge which points at $\msf{trans}_w([\inft \mtt{a}]_{w}^\mrm{R},2,\inft)={F}([\inft\inft\mtt{a}3]^\mrm{R}_w) = [\inft \mtt{a} \inft]_w^\mrm{R}$.
			On the other hand, the 2-edge of $[\inft\inft \mtt{a}]_{w}^\mrm{R}$ is simply copied for $[\inft \texttt{a}]_{wa}^\mrm{R}$.
		}
\end{figure}
\begin{lemma}[Outgoing edges of the LRS node]\label{lem:construction_outgoing_edges}
	For $u \in V_w \cap V_{wa}$, 
	\begin{itemize}
		\item $([y]^\mrm{R}_{wa},b,u) \in E_{wa}$ if and only if\/ $([y]^\mrm{R}_{w},b,u) \in E_{w}$,
		\item $([x]^\mrm{R}_{wa},b,u) \in E_{wa}$ if and only if\/ $\msf{trans}_w([y]^\mrm{R}_{w},|x|,b) = u$ if and only if
			either $([y]^\mrm{R}_{w},b,u) \in E_{w}$ and $\Zinf{b}{|x|} \neq \inft$
			or\/ $\msf{trans}_w([y]^\mrm{R}_{w},|x|,\inft) = u$ and $\Zinf{b}{|x|} = \inft$.
	\end{itemize}
	Moreover,
	\begin{itemize}
		\item $([x]^\mrm{R}_{wa},\Zinf{a}{|x|},\rec{wa}{wa}) \in E_{wa}$ if and only if
		 $([y]^\mrm{R}_{wa},\Zinf{a}{|y|},\rec{wa}{wa}) \in E_{wa}$.
		 \item If $([x]^\mrm{R}_{wa},\inft,\rec{wa}{wa}) \in E_{wa}$, then $Z = \{\, i \in \calN \cap \msf{Children}_w(\rec{y}{w}) \mid i > |x|\,\}$ is empty.
	\end{itemize}
\end{lemma}
\begin{proof}
	The claims on edges toward nodes $u \in V_w$ is an immediate consequence of the definition of PDAWG edges and the soundness of the function $\msf{trans}$ (Algorithm~\ref{alg:pdawgtransition} and Lemma~\ref{lem:pdawgmatching}).
	We prove the third and fourth claims.
	
	Suppose $([x]^\mrm{R}_{wa},\Zinf{a}{|x|},\rec{wa}{wa}) \in E_{wa}$, i.e., $\msf{RPos}_{wa}(\lrangle{xa}) = \{|wa|\}$.
	Then, $x \in \PSuffix(w)$ and $x \equiv_w^\mrm{R} y$ imply $y \in \PSuffix(w)$. 
	Hence, $\lrangle{ya} \in \PSuffix(wa)$.
	Since $\msf{RPos}_{wa}(\lrangle{ya}) \subseteq \msf{RPos}_{wa}(\lrangle{xa})$, we have $\msf{RPos}_{wa}(\lrangle{ya}) = \{|wa|\}$ and thus $([y]^\mrm{R}_{wa},\Zinf{a}{|y|},\rec{wa}{wa}) \in E_{wa}$.
	Conversely, suppose $([y]^\mrm{R}_{wa},\Zinf{a}{|y|},\rec{wa}{wa}) \in E_{wa}$, i.e., $\msf{RPos}_{wa}(\lrangle{ya}) = \{|wa|\}$.
	We have $\lrangle{xa} \in \PSuffix(\lrangle{ya}) \subseteq \PSuffix(wa)$.
	Since $\lrangle{xa}$ is longer than the LRS $x$, it cannot occur in $w$.
	Thus, $|wa|$ is the only end position of $\lrangle{xa}$ in $wa$.
	Therefore, $(\rec{x}{wa},\Zinf{a}{|x|},\rec{wa}{wa}) \in E_{wa}$.
	
	Suppose $(\rec{x}{wa},\inft,\rec{wa}{wa}) \in E_{wa}$ and $Z \neq \emptyset$.
	The fact $(\rec{x}{wa},\inft,\rec{wa}{wa}) \in E_{wa}$ implies $x \inft \in \PSuffix(wa)$.
	The fact $Z \neq \emptyset$ implies, by Lemma~\ref{lem:pdawgmatching}, $x \inft \in \PFactor(w)$, which contradicts that $x$ is the LRS.
\end{proof}

Let us turn our attention to the incoming edges of the LRS node.
Lemma~\ref{lem:construction_incoming_edges} is no more than a direct implication of the definition of edges of PDAWGs.
If a node $u$ has an outgoing edge pointing at the LRS node, the edge will point at $[y]^\mrm{R}_{wa}$ in $\msf{PDAWG}(wa)$ if and only if $\longest{u}$ is longer than the pre-LRS $x'$.
Otherwise, it will point at $[x]^\mrm{R}_{wa}$ possibly changing the label to $\inft$ if necessary.
\begin{lemma}[Incoming edges of the LRS node]\label{lem:construction_incoming_edges}
	We have
	\begin{itemize}
		\item $(u,b,[y]^\mrm{R}_{wa}) \in E_{wa}$ if and only if\/ $(u,b,[y]^\mrm{R}_{w}) \in E_{w}$ and $|\longest{u}| > |x'|$,
		\item $(u,b,[x]^\mrm{R}_{wa}) \in E_{wa}$ if and only if\/ $(u,b,[y]^\mrm{R}_{w}) \in E_{w}$ and $|\longest{u}| \le |x'|$, where $b=\Zinf{a}{|\longest{u}|}$ and $\longest{u} \in \PSuffix(x')$.
	\end{itemize}
\end{lemma}
\begin{proof}
	Recall that $(u,b,[y]^\mrm{R}_{w}) \in E_{w}$ if and only if $\longest{u}b \in [y]^\mrm{R}_{w} = [x]^\mrm{R}_{wa} \cup [y]^\mrm{R}_{wa}$.
	Here, for each $(u,b,[y]^\mrm{R}_{w}) \in E_{w}$, clearly $\longest{u}b \in [x]^\mrm{R}_{wa}$ if and only if $\longest{u} \le |x'|$.
	In this case, $\longest{u}b \in [x'a']^\mrm{R}_{wa}$ implies $b=\Zinf{a}{|\longest{u}|}$ and $\longest{u}b \in \PSuffix(x'a')$, so $\longest{u} \in \PSuffix(x')$.
\end{proof}
Therefore, the incoming edges of $\rec{x}{wa}$ can be constructed by visiting nodes $u$ on the suffix link chain of the pre-LRS node and checking their $\Zinf{a}{|\longest{u}|}$-edges.
Note that in the online construction of a DAWG, the edge from the pre-LRS node $[x']^\mrm{R}_{w}$ to the LRS node $[y]^\mrm{R}_{w}$ in the old DAWG will be inherited to the new node $[x]^\mrm{R}_{wa}$ in the new DAWG, since always $\longest{\rec{x'}{w}} = x'$ holds.
However, it is not necessarily the case in the PDAWG construction.
\begin{example}
Let us come back to the example in Figure~\ref{fig:incoming_edge},
 where $x=\mtt{a}\inft$, $x'=\mtt{a}$, and $y=\inft\mtt{a}2$ for $wa=\inft 2 \mtt{a}\inft\mtt{a}\inft$.
The outgoing edge of the LRS node $\rec{x}{w} \in V_w$ is inherited to both $\rec{x}{wa}$ and $\rec{y}{w}$ in $\PDAWG(wa)$.

The LRS node has one incoming edge in $\PDAWG(w)$.
Since $|\longest{\rec{x'}{w}}| = |\inft \mtt{a}| > |x'|=|\mtt{a}|$,
the 2-edge from $[x']^\mrm{R}_{w}$ to $[y]^\mrm{R}_{w}$ in $\msf{PDAWG}(w)$ is kept as the 2-edge from $[x']^\mrm{R}_{wa}$ to $[y]^\mrm{R}_{wa}$ in $\msf{PDAWG}(wa)$ and, as a result, the new node $[x]^\mrm{R}_{wa}$ has no incoming edges.

Recall that in Figure~\ref{fig:ltor_example}, we have $x=\mtt{a}$, $x'=\varepsilon$, and $y=\inft\mtt{a}$ for $wa=\inft \mtt{a}2\mtt{a}\inft\mtt{aa}$.
The LRS node has two outgoing edges: $\inft$-edge and $2$-edge.
The new node $[y]^\mrm{R}_{wa} \in V_{wa}$ inherits those two.
In addition, $[y]^\mrm{R}_{wa}$ gets an $\mtt{a}$-edge pointing at the new sink $\rec{wa}{wa}$.
On the other hand, $[x]^\mrm{R}_{wa} \in V_{wa}$ has the $\inft$-edge pointing at the node $\msf{trans}_w(\rec{x}{w},|x|,\inft)$, but it has no $2$-edge.
In addition, $[x]^\mrm{R}_{wa}$ gets an $\mtt{a}$-edge toward $\rec{wa}{wa}$, too.

The LRS node has two incoming edges labeled with $\mtt{a}$.
The one from $\rec{\varepsilon}{w}$ is inherited to $\rec{x}{wa}$ by $|\varepsilon| \le |x'| = 0$,
and the one from $\rec{\inft}{w}$ is inherited to $\rec{y}{wa}$ by $|\inft|=1 > |x'|$.
\end{example}

Updates of suffix links simply follow the definition.
\begin{lemma}[Suffix link update]\label{lem:construction_suffixlinks}
	Suppose $V_{wa}=V_w \cup \{[wa]^\mrm{R}_{wa}\}$, i.e., no node split happens.
	Then, for each $u \in V_{wa}$,
	\[
		{F}_{wa}(u) = \begin{cases}
			[\msf{LRS}(wa)]^\mrm{R}_{wa}	& \text{if\/ $u = [wa]^\mrm{R}_{wa}$,}
\\			{F}_{w}(u) 		& \text{otherwise.}
		\end{cases}
	\]
	Suppose $[x]^\mrm{R}_w \neq [x]^\mrm{R}_{wa}$ for $x=\msf{LRS}(wa)$, where
	 $V_{wa}=V_w \setminus \{[y]^\mrm{R}_{w}\} \cup \{[wa]^\mrm{R}_{wa},[ x ]^\mrm{R}_{wa},\linebreak[1][ y ]^\mrm{R}_{wa}\}$ for $y = \longest{[x]^\mrm{R}_{w}}$.
	Then, for each $u \in V_{wa}$,
	\[
		{F}_{wa}(u) = \begin{cases}
			[x]^\mrm{R}_{wa}	& \text{if\/ $u \in \{ [wa]^\mrm{R}_{wa} , [y]^\mrm{R}_{wa}\}$,}
\\			{F}_{w}([y]^\mrm{R}_{w})		& \text{if\/ $u = [x]^\mrm{R}_{wa}$,}
\\			{F}_{w}(u) 	& \text{otherwise.}
		\end{cases}
	\]
\end{lemma}
\begin{proof}
	Recall that $\msf{LRS}(wa)$ is the longest p-suffix of $wa$ that occurs in $w$.
	That is, all the pv-suffixes of $wa$ longer than $\msf{LRS}(wa)$ belongs to the sink $\rec{wa}{wa}$.
	Therefore, in both cases, $F(\rec{wa}{wa}) = \rec{x}{wa}$.
	If $[x]^\mrm{R}_w = [x]^\mrm{R}_{wa}$, all the suffix links in $\PDAWG(w)$ are kept.
	If $[x]^\mrm{R}_w \neq [x]^\mrm{R}_{wa}$, $F(\rec{y}{wa}) = [x]^\mrm{R}_{wa}$, since the longest p-suffix of $y$ not in $\rec{y}{wa}$ is $x$.
	On the other hand, the longest p-suffix of $x$ not in $\rec{x}{wa}$ is the longest p-suffix of $y$ not in $\rec{y}{w}$, i.e, $F_w(\rec{y}{w})$.
\end{proof}

\begin{algorithm2e}[t!]
	\caption{Constructing $\PDAWG(T)$\label{alg:construction}}
	\SetVlineSkip{0.5mm}
	Let $V \leftarrow \{\, \top,\rho \,\}$, $E \leftarrow \{\,(\top,a,\rho) \mid a \in \Sigma \cup \{\inft\}\,\}$, ${F}(\top) \leftarrow \top$,  ${F}(\rho) \leftarrow \top$, $\msf{len}(\top)=-1$, $\msf{len}(\rho) \leftarrow 0$,  $\mathit{sink} \leftarrow \rho$, and $t \leftarrow \lrangle{T}$\;
	\For{$i \leftarrow 1$ to $|t|$}{%
		Let $a \leftarrow t[i]$ and $u \leftarrow \mathit{sink}$\;
		Create a new node and let $\mathit{sink}$ be that node with $\msf{len}(\mathit{sink})=i$\;
		\While{$\msf{trans}(u,\msf{len}({F}(u))+1,\Zinf{a}{\msf{len}({F}(u))+1}) = \msf{Null}$\label{algln:while}}{%
			Let $\msf{child}(u,\Zinf{a}{\msf{len}(u)}) \leftarrow \mathit{sink}$ and $u \leftarrow {F}(u)$\;
		}
		\tcp{$u$ corresponds to $[\msf{preLRS}(t[:i])]^\mrm{R}_{t[:i-1]}$}
		\If(\tcp*[h]{\scalebox{0.98}[1]{$\msf{preLRS}(t[:i]) = \longest{[\msf{preLRS}(t[:i])]^\mrm{R}_{t[:i-1]}}$}}){\label{algln:length1}%
			$\Zinf{a}{\msf{len}(u)} \in \msf{Children}(u)$}{%
			Let $k \leftarrow \msf{len}(u) + 1$ and $v \leftarrow \msf{child}(u,\Zinf{a}{\msf{len}(u)})$%
		}\Else(\tcp*[h]{$\msf{preLRS}(t[:i]) \neq \longest{[\msf{preLRS}(t[:i])]^\mrm{R}_{t[:i-1]}}$}){%
			Let $k \leftarrow \min \{a,\,\max (\msf{Children}(u) \cap \calN ) \}$,
			$v \leftarrow \msf{trans}(u,k-1,\inft)$,
			$\msf{child}(u,\Zinf{a}{\msf{len}(u)}) \leftarrow \mathit{sink}$,
			and $u \leftarrow {F}(u)$\;
			\tcp{$u$ corresponds to $F_{[t[:i-1]]}([\msf{preLRS}(t[:i])]^\mrm{R}_{t[:i-1]})$}
		}
		\tcp{$v$ corresponds to $[\msf{LRS}(t[:i])]^\mrm{R}_{t[:i-1]}$ and $k = |\msf{LRS}(t[:i])|$}\label{algln:sinkedges}
		\lIf(\tcp*[h]{No node split}){$\msf{len}(v) = k$}{%
			Let ${F}(\mathit{sink}) \leftarrow v$\label{algln:nosplit}%
		}\Else(\tcp*[h]{Node split}){
			Create a new node $v'$;
			\tcp*[h]{$v'$ corresponds to $[\msf{LRS}(t[:i])]^\mrm{R}_{t[:i]}$}\\
			Let $\msf{len}(v') \leftarrow k$\;
			\tcp*[h]{Outgoing edges of the new node}\\
			\For{each $b \in \msf{Children}(v)$ such that $\Zinf{b}{k} \neq \inft$\label{algln:outgoing1}}{%
				{Let $\msf{child}(v',b) \leftarrow \msf{child}(v,b)$\;}
			}
			\lIf{$\msf{trans}(v,k,\inft) \neq \mathsf{Null}$}{Let $\msf{child}(v',\inft) \leftarrow \msf{trans}(v,k,\inft)$}\label{algln:alledges}
			\tcp*[h]{Incoming edges of the new node}\\
			\While{$\msf{child}(u,\Zinf{a}{\msf{len}(u)}) = v$\label{algln:incoming}}{%
				Let $\msf{child}(u,\Zinf{a}{\msf{len}(u)}) \leftarrow v'$ and $u \leftarrow {F}(u)$\;
			}\label{algln:incoming2}
			\tcp*[h]{Suffix links}\\
			Let ${F}(v') \leftarrow {F}(v)$, ${F}(v) \leftarrow v'$ and ${F}(\mathit{sink}) \leftarrow v'$;
		}
	}
	\textbf{return} $(V,E,{F})$\;
\end{algorithm2e}

Algorithm~\ref{alg:construction} constructs PDAWGs based on the above lemmas, 
where $\rho$ and $t[:i]$ in the pseudo code corresponds to the source node $\{\varepsilon\}$ and $wa$ in the main body of the text, respectively. 
Figure~\ref{fig:ltor_example} illustrates an example run.
For technical convenience, like the standard DAWG construction algorithm, we add a dummy node $\top$ to the PDAWG that has edges to the source node labeled with all elements of $\Sigma \cup \{\inft\}$ and let ${F}(\rho)={F}(\top)=\top$.
This trick allows us to uniformly treat the special case where the LRS node is $\rho$, in which case $\top$ is regarded as the pre-LRS node.
Each node $u$ does not remember the elements of $u$ but it remembers $\msf{len}(u)= |\longest{u}|$.
For the dummy node $\top$, we let $\msf{len}(\top)=-1$.
Note that $|\shortest{u}| = |\msf{len}({F}(u))|+1$ (used in Line~\ref{algln:while}).
Hereafter, we use functions ${F}$, $\msf{child}$, $\msf{trans}$, etc.\ without a subscript specifying a text, to refer to the data structure that the algorithm is manipulating, rather than the mathematical notion relative to the text.
Of course, we design our algorithm so that those functions coincide with the corresponding mathematical notions.

Suppose we have constructed $\msf{PDAWG}(w)$ and want to obtain $\msf{PDAWG}(wa)$.
The sink node of $\msf{PDAWG}(w)$, which we denote as $\mathit{oldsink}$, corresponds to $[w]^\mrm{R}_w$.
We first make a new sink node $\mathit{newsink} =[wa]^\mrm{R}_{wa}$ and let $\msf{len}(\mathit{newsink})=|wa|$.
Then, in the \textbf{while} loop of Line~\ref{algln:while}, we visit $u_0,u_1,\dots,u_j$ on the suffix link chain of $\mathit{oldsink}$, until we find the pre-LRS node $u_j = [\msf{preLRS}(wa)]^\mrm{R}_w$.
By Lemma~\ref{lem:prelrs}.1, we identify $u_j$ when exiting the \textbf{while} loop.
For each node $u_i$ with $i < j$, we make an edge labeled with $\Zinf{a}{\msf{len}(u_i)}$ pointing at $\mathit{newsink}$ by Lemma~\ref{lem:construction_sink_edges}.
The algorithm identifies the length $k$ of the LRS in Lines~\ref{algln:length1}--\ref{algln:sinkedges} based on Lemma~\ref{lem:prelrs}.2.
If $k-1 < \msf{len}(u_j)$, the pre-LRS node $u_j$ also has an edge pointing at $\mathit{newsink}$ by Lemma~\ref{lem:construction_sink_edges}.
This is done at Line~\ref{algln:sinkedges}.
At this moment, we have created all the incoming edges of $\mathit{newsink}$, possibly except the one from $\rec{\msf{preLRS}(wa)}{wa}$, if necessary, because that node $\rec{\msf{preLRS}(wa)}{wa}$ has not yet been created in the case of node split.
We then reach the LRS node $v = [\msf{LRS}(wa)]^\mrm{R}_w = \msf{trans}(u_j,k-1,\Zinf{a}{k-1})$ by Lemma~\ref{lem:prelrs}.3.
At the moment entering Line~\ref{algln:nosplit}, the variable $u$ represents the first node on the suffix link chain of the pre-LRS node such that $\longest{u}$ is not longer than the pre-LRS.
It is just the pre-LRS node $u_j$ if $k-1 = \msf{len}(u_j)$, and it is $F(u_j)$ otherwise.
Then, in both cases, $u$ is the first node whose edge toward the LRS node will be redirected to $\rec{x}{wa}$ in the \textbf{while} loop of Line~\ref{algln:incoming} in accordance with Lemma~\ref{lem:construction_incoming_edges}, if it has one.

We compare $k$ and $\msf{len}(v)$ to decide whether the LRS node should be split based on Lemma~\ref{lem:prelrs}.4.
If $|\msf{LRS}(wa)| = \msf{len}(v)$, the node $v$ will not be split, in which case we obtain $\msf{PDAWG}(wa)$ by making ${F}(\mathit{newsink}) = v$ at Line~\ref{algln:nosplit} by Lemma~\ref{lem:construction_suffixlinks}.
Suppose $k < \msf{len}(v)$.
In this case, the LRS node $v$ must be split.
We reuse the old node $v$, which used to correspond to $[\msf{LRS}(wa)]^\mrm{R}_w$, as a new node corresponding to $[\longest{[\msf{LRS}(wa)]^\mrm{R}_w}]^\mrm{R}_{wa}$, and create another new node $v'$ for $[\msf{LRS}(wa)]^\mrm{R}_{wa}$ with $\msf{len}(v')=k$.
Edges are determined in accordance with Lemmas~\ref{lem:construction_outgoing_edges} and~\ref{lem:construction_incoming_edges}.
The outgoing edges from $v$ shall be kept.
We create outgoing edges of $v'$ referring to the corresponding transitions from $v$.
If $(v,b,u) \in E$ with $\Zinf{b}{k} \neq \inft$, then we add $(v',b,u)$ to $E$ in the \textbf{for} loop of Line~\ref{algln:outgoing1}.
In addition, we add $(v',\inft,\msf{trans}(v,k,\inft))$ to $E$ if $\msf{trans}(v,k,\inft) \neq \msf{Null}$ at Line~\ref{algln:alledges}.
We note that thanks to the third and fourth claims of Lemma~\ref{lem:construction_outgoing_edges}, this correctly creates an edge $(v',\inft,\mathit{newsink})$ if necessary.
We should have $(v',\inft,\mathit{newsink}) \in E$ if and only if $(v,\Zinf{a}{\msf{len}(v)},\mathit{newsink}) \in E \wedge \Zinf{a}{k}=\inft$ if and only if $\msf{trans}(v,k,\Zinf{a}{k}) = \mathit{newsink}$.
All incoming edges of $v$ from nodes on the suffix link chain of $u$ in $\msf{PDAWG}(w)$ are redirected to $v'$ in the \textbf{while} loop of Line~\ref{algln:incoming}, where $u$ is the pre-LRS node $u_j$ if $\longest{u_j}$ is the pre-LRS, and it is $F(u_j)$ otherwise.

At last, the suffix link from $\mathit{newsink}$ to $v$ and the suffix link from $v$ to $v'$ are determined in accordance with Lemma~\ref{lem:construction_suffixlinks}.

\begin{remark}
	One can compute $\ell_u = \min\msf{RPos}(u)$ online simply by letting $\ell_{[wa]^\mrm{R}_{wa}} = |wa|$ when creating the new sink node $[wa]^\mrm{R}_{wa}$.
	When the LRS node $[y]_{w}^{\mrm{R}}$ is split into $[y]_{wa}^{\mrm{R}}$ and $[x]_{wa}^{\mrm{R}}$, we have $\min\msf{RPos}([y]_{w}^{\mrm{R}})=\min\msf{RPos}([y]_{wa}^{\mrm{R}})=\min\msf{RPos}([x]_{wa}^{\mrm{R}})$.
	So, it is enough to copy the $\ell$ value.
\end{remark}

\subsection{Time complexity analysis}
Let us call an edge $(u,a,v)$ \emph{primary} if $\longest{v} = \longest{u} \cdot a$, and \emph{secondary} otherwise.
The following lemma is an adaptation of the corresponding one for DAWGs by Blumer et al.~\cite{DAWG}.
\begin{lemma}\label{lem:suffixlinkchain}
	Let $\msf{SC}_w(u)$ be the set of nodes on the suffix link chain of a node $u$.
	If\/ $\msf{PDAWG}(w)$ has a primary edge from $u$ to $v$,
	then the total number of secondary edges from nodes in $\msf{SC}_w(u)$ to nodes in $\msf{SC}_w(v) $ is bounded by $|\msf{SC}_w(u)| - |\msf{SC}_w(v)| + |\Pi| + 1$.
\end{lemma}
\begin{proof}
	Let us count the number of edges from nodes in $\msf{SC}_w(u)$ to $\msf{SC}_w(v)$.
	Baker~\cite[Lemma~1]{PMA} showed that in a parameterized suffix tree, each path from the root to a leaf has at most $|\Pi|$ nodes with \emph{bad suffix links}.
	Through the duality of PDAWGs and parameterized suffix trees stated in Lemma~\ref{lem:duality}, this means that
	$\msf{SC}_w(v)$ contains at most $|\Pi| + 1$ nodes which have no incoming primary edges, where the additional one node is the root of the PDAWG.
	Since each node has at most one incoming primary edge, the number of primary edges in concern is at least $|\msf{SC}_w(v)| - |\Pi| - 1$ in total.
	Since each node in $\msf{SC}_w(u)$ has just one outgoing edge to $\msf{SC}_w(v)$, we obtain the lemma.
\end{proof}
\begin{theorem} \label{theo:pdawg_online}
	Given a string $T$ of length $n$, Algorithm~\ref{alg:construction} constructs $\PDAWG(T)$ in $O(n |\Pi| \log(|\Pi|+|\Sigma|))$ time and $O(n)$ space online, by reading $T$ from left to right.
\end{theorem}
\begin{proof}
	Since the size of a PDAWG is bounded by $O(n)$ (Theorem~\ref{thm:pdawg_size}) and nodes are monotonically added, it is enough to bound the number of edges and suffix links that are deleted.
	In each iteration of the \textbf{for} loop, at most one suffix link is deleted. So at most $n$ suffix links are deleted in total.
	We count the number of edges whose target is altered from $v = [\msf{LSR}(wa)]^\mrm{R}_w$ to $v' = [\msf{LSR}(wa)]^\mrm{R}_{wa}$ on Line~\ref{algln:incoming} when updating $\msf{PDAWG}(w)$ to $\msf{PDAWG}(wa)$.
	Let $k_i$ be the number of such edges at the $i$-th iteration of the \textbf{for} loop.
	Note that those are all secondary edges from a node in $\msf{SC}_w(u_0)$ for the pre-LRS node $u_0$.
	By Lemma~\ref{lem:suffixlinkchain},
	\begin{align*}
	\sum_{i=1}^n k_i & \le \sum_{i=1}^n \big( |\msf{SC}_{wa}(w)|-|\msf{SC}_{wa}(wa)|+ |\Pi| +1  \big)
\\		& \le \sum_{i=1}^n \big( |\msf{SC}_{w}(w)|-|\msf{SC}_{wa}(wa)|+ |\Pi| +1 \big)
\\		&= |\msf{SC}_{\varepsilon}(\varepsilon)|-|\msf{SC}_{t}(t)| + (|\Pi| +1)n \in {O}(|\Pi| n)\,.
\tag*{\qedhere}	\end{align*}
\end{proof}

Since the suffix links of $\PDAWG(T)$ form the p-suffix tree of $\rev{T}$
(see Subsection~\ref{sec:pdawg_pst}),
the following corollary is immediate from Theorem~\ref{theo:pdawg_online}.
\begin{corollary}
  The p-suffix tree of a string $S$ of length $n$
  can be constructed 
  in $O(n |\Pi| \log(|\Pi|+|\Sigma|))$ time and $O(n)$ space online, by reading $S$ from right to left.
\end{corollary}
Differently from the online DAWG construction algorithm~\cite{DAWG}, we have the factor $|\Pi|$ in our algorithm complexity analysis.
Actually, our algorithm takes time proportional to the difference of the old and new PDAWGs modulo logarithmic factors, as long as the difference is defined so that the split node $[\msf{LRS}(wa)]^\mrm{R}_w$ automatically becomes $[\longest{[\msf{LRS}(wa)]^\mrm{R}_w}]^\mrm{R}_{wa}$ rather than $[\msf{LRS}(wa)]^\mrm{R}_{wa}$.
In this sense, our algorithm is optimal.
It is open whether we could improve the analysis.

\section{Concluding remarks}\label{sec:conclusion}

In this paper, we proposed a new indexing structure for
parameterized pattern matching---the PDAWGs.
We showed that $\PDAWG(T)$ for an input text string $T$ of length $n$
over a static alphabet $\Sigma$ and a parameterized alphabet $\Pi$
can be built in $O(n|\Pi|\log (|\Pi+|\Sigma||))$ time
with $O(n)$ space, in a right-to-left online manner.
The duality of our PDAWGs and parameterized suffix trees~\cite{Baker93}
permits us $O(n)$-time offline construction of $\PDAWG(T)$
provided that the p-suffix tree of $\rev{T}$ has already been built.
It also gives us a linear-space bidirectional index for parameterized pattern matching
in $O(m\log(|\Pi|+|\Sigma|) + \occ)$ time.

The major open question is whether our upper bound
$O(n|\Pi|\log (|\Pi+|\Sigma||))$ for the
time complexity of online construction of the PDAWGs is tight.
Our construction algorithm is ``optimal'' in the sense that,
given a new character $a$ to append,
the time for updating $\PDAWG(T)$ to $\PDAWG(Ta)$
is proportional to the difference of the two DAWGs,
ignoring the logarithmic factors which are the costs for searching for edges and/or suffix links.
We have not found an instance that requires $\Omega(n|\Pi|)$ changes
into the DAWG during the whole online construction.

\subsection*{Acknowledgment}
The authors are deeply grateful to the anonymous reviewer for many valuable comments that have improved the paper's readability and preciseness.
This work was supported by JSPS KAKENHI Grant Numbers
JP19K20208, 
JP18K18002, 
JP18H04091, JP18K11150, 
JP17H01697, 
JP16H02783, JP20H04141, 
JP15H05706, JP21K11745, 
JP18H04098, 
and
JST PRESTO Grant Number JPMJPR1922. 

\bibliography{ref}

\end{document}